\documentclass[printer]{aa}
\usepackage{graphicx}

\usepackage{natbib}
\bibpunct{(}{)}{;}{a}{}{,} 

\usepackage{txfonts}
\usepackage{longtable}
\usepackage{listings}
\usepackage{color}
\usepackage{textcomp}

\begin{document}

\title{
Characterising open clusters in the solar neighbourhood with the Tycho-\textit{Gaia} Astrometric Solution
}

   \subtitle{}

\author{
T. Cantat-Gaudin\inst{\ref{OAPD}}
\and
A. Vallenari\inst{\ref{OAPD}}
\and
R. Sordo\inst{\ref{OAPD}}
\and
F. Pensabene\inst{\ref{OAPD},\ref{UNIPD}}
\and
A. Krone-Martins\inst{\ref{SIMUL}}
\and
A. Moitinho\inst{\ref{SIMUL}}
\and
C. Jordi\inst{\ref{IEECUB}}
\and
L. Casamiquela\inst{\ref{IEECUB}}
\and
L. Balaguer-N{\'u}nez\inst{\ref{IEECUB}}
\and
C. Soubiran\inst{\ref{LAB}}
\and
N. Brouillet\inst{\ref{LAB}}
}

\institute{
INAF-Osservatorio Astronomico di Padova, vicolo Osservatorio 5, 35122 Padova, Italy\label{OAPD}
\and
Dipartimento di Fisica e Astronomia, Universit\`a di Padova, vicolo Osservatorio 3, 35122 Padova, Italy\label{UNIPD}
\and
SIM, Faculdade de Ci\^encias, Universidade de Lisboa, Ed. C8, Campo Grande, P-1749-016 Lisboa, Portugal\label{SIMUL}
\and
Institut de Ci\`encies del Cosmos, Universitat de Barcelona (IEEC-UB), Mart\'i i Franqu\`es 1, E-08028 Barcelona, Spain\label{IEECUB}
\and
Laboratoire d’Astrophysique de Bordeaux, Univ. Bordeaux, CNRS, UMR 5804, 33615 Pessac, France\label{LAB}
}

\date{Received date / Accepted date }

\abstract
{The Tycho-\textit{Gaia} Astrometric Solution (TGAS) subset of the first \textit{Gaia} catalogue contains an unprecedented sample of proper motions and parallaxes for two million stars brighter than $G\sim12$\,mag. }
{We take advantage of the full astrometric solution available for those stars to identify the members of known open clusters and compute mean cluster parameters using either TGAS or the fourth U.S. Naval Observatory CCD Astrograph Catalog (UCAC4) proper motions, and TGAS parallaxes.}
{We apply an unsupervised membership assignment procedure to select high probability cluster members, we use a Bayesian/Markov Chain Monte Carlo technique to fit stellar isochrones to the observed 2MASS $JHK_S$ magnitudes of the member stars and derive cluster parameters (age, metallicity, extinction, distance modulus), and we combine TGAS data with spectroscopic radial velocities to compute full Galactic orbits.}
{We obtain mean astrometric parameters (proper motions and parallaxes) for 128 clusters closer than about 2\,kpc, and cluster parameters from isochrone fitting for 26 of them located within a distance of 1\,kpc from the Sun. We show the orbital parameters obtained from integrating 36 orbits in a Galactic potential.}
{}


\maketitle{}

\section{Introduction}

Open clusters (OCs) are basic constituents of our Galaxy. It is believed that most stars, if not all, are born within clusters, which are later disrupted and their population mixes with what is referred to as the ``field population'' \citep[see e.g.][]{Lada03}. The homogeneous age and chemical composition of stars within a cluster makes them valuable tracers of the properties of the Galactic disk \citep{CantatGaudin14m11,Vasquez10}, as well as testbeds for stellar evolution models.
OCs trace the radial metallicity gradient \citep{CantatGaudin16,Jacobson16}; their age, metallicity, and kinematics are important to define how the radial migration process can affect the disk \citep{Minchev17}.

Although different authors may use slightly different definitions for what constitutes an OC, it is generally considered that stars within a cluster have a common origin (they have similar ages and chemical compositions) and are bound by gravity (they are physically close to each other and share a common orbit through the disk of the Milky Way). Historically, many stellar clusters have been serendipitously discovered as local over-densities of stars \citep[e.g.][]{Messier1781}. 

The majority of known OCs are located within 2\,kpc of the Sun. The most recent version of the OC catalogue of \citet{Dias02} (hereafter DAML) quotes about 2200 objects, while the Milky Way Star Clusters catalogue of \citet{Kharchenko13} lists over 3000 objects, many of which are putative or candidate clusters. On one hand, the cluster census is far from being complete, especially at the faint end where small and sparse objects and remnants of disrupted clusters can escape detection \citep{Bica11} (even when their stars are visible) because they do not stand as significant over-densities. On the other hand, cluster searches produce false positives and astrometric information is necessary to distinguish a genuine cluster (sharing common proper motions and parallaxes) from a coincidental asterism.

{Hipparcos \citep{Perryman97} was the first space mission dedicated to astrometry. The data it collected yielded parallaxes for about 120,000 stars in our Galaxy \citep{ESA97,vanLeeuwen07} and allowed for the construction of the 2.5 million stars proper motion catalogue Tycho-2 \citep{Hog00}. The ongoing \textit{Gaia} space mission \citep{Perryman01,Gaia16} will deliver proper motions and parallaxes for a billion sources, dwarfing the Hipparcos catalogue by four orders of magnitude. \textit{Gaia} is expected to revolutionize the field, discovering new objects \citep{Koposov17} and providing accurate astrometrically based characterizations of thousands of clusters. The \textit{Gaia} mission first data release \citep[hereafter GDR1][]{GaiaDR1} dataset, which contains positions and $G$-band magnitudes for a billion sources, also contains a subset called the Tycho-Gaia Astrometric Solution (TGAS), consisting of two million sources with a full astrometric solution obtained through the combination of \textit{Gaia} observations and the positions listed in Hipparcos and Tycho-2 \citep{Lindegren16}. The next \textit{Gaia} data releases are expected to deliver astrometric measurements for more than a billion stars. Handling such large, multi-dimensional datasets (whose applications extend well beyond stellar cluster science) requires automated methods in order to identify and select cluster stars and to characterise stellar clusters \citep[see e.g.][with Hipparcos data]{Robichon99}. In this paper we chose to follow the unsupervised photometric membership assignment in stellar clusters approach (UPMASK) introduced by \citet{KroneMartins14}, applying it to astrometric data for the first time.

The aims of this paper are two-fold: i) to validate the use of tools that can be applied to multi-dimensional datasets such as the two million sources in TGAS, or the billion sources of the upcoming further \textit{Gaia} releases, and ii) to update the cluster census in the solar neighbourhood, deriving memberships, and mean parallaxes and proper motions from the stellar clusters that can be clearly identified in the TGAS catalogue.

In Sect.~\ref{sec:membership} we present the method we applied to select probable cluster members in the TGAS catalogue for 128 OCs, extending the use of the code UPMASK \citep{KroneMartins14} to astrometric data.
Section \ref{sec:meanparams} presents the mean astrometric parameters we computed for these clusters, and
Sect.~\ref{sec:base9} presents the ages and cluster parameters we obtained for 34 of them by isochrone fitting using the Bayesian approach of the code \citep[BASE-9][]{vonHippel06}.
Finally, Sect.~\ref{sec:discussion} discusses the results obtained in the context of the Milky Way disk, and Sect.~\ref{sec:conclusion} closes with concluding remarks.

\section{Membership from TGAS data} \label{sec:membership}

In this section, we compare the uncertainties in the proper motions in the TGAS and UCAC4 datasets, and discuss the membership of the clusters present in TGAS.

\subsection{Proper motions in the TGAS dataset} \label{sec:tgas}
In this paper the parallaxes used were always the TGAS parallaxes, but the proper motions were either taken from TGAS or from UCAC4 \citep{Zacharias12}.
Because the \textit{Gaia} scanning law in the first 14 months that are included in GDR1 has been favouring more frequent visits to regions closer to the ecliptic poles \citep{GaiaDR1}, the quality of the TGAS data varies significantly between regions across the sky. In some areas the uncertainties on the proper motions are larger than those of the UCAC4 catalogue. 
The uncertainties on proper motions of TGAS stars are in the range 0.5--2.6\,mas\,yr$^{-1}$, with a median value of 1.1 \,mas\,yr$^{-1}$ \citep{Lindegren16}. UCAC4 proper motion formal uncertainties are in the range 1--10 \,mas\,yr$^{-1}$, with possible systematics of the order of 1--4 \,mas\,yr$^{-1}$ \citep{Zacharias13}, on average higher than TGAS uncertainties. However, the effect of the \textit{Gaia} scanning law is that in some regions TGAS uncertainties can be on average higher than in UCAC4. This effect is clearly visible in Fig.~\ref{fig:mollweide_map_error_ratio}, which shows a comparison between the proper motion uncertainties in UCAC4 and TGAS for the stars present in both catalogues.
Figure~\ref{fig:map_PMdiff_TGAS_UCAC4} shows the mean proper motion difference between TGAS and UCAC4 for those stars. Local systematic differences of the order of 1 to 3\,mas\,yr$^{-1}$ are present all across the sky. 

In each field under analysis, in order to see the cluster members as a more compact group in astrometric space, we used the proper motion catalogue, which provided the smallest median uncertainty, thus using TGAS proper motions in the regions where the ratio shown in Fig.~\ref{fig:mollweide_map_error_ratio} is above one (e.g. for NGC~2360, for which the TGAS and UCAC4 proper motions are shown in Fig.~\ref{fig:ngc2360_pmcorrelations}). 
In practice, whether UCAC4 or TGAS have better proper motion errors for individual stars does not depend solely on the region, but also on the type of astrometric prior used in TGAS. Even in the regions where UCAC4 provides better median errors than TGAS, the subset of TGAS stars whose proper motion was determined using their \textsc{Hipparcos} positions as astrometric prior always has smaller errors than UCAC4 (typically under 0.1\,mas). Those stars only account for 5\% of the whole TGAS dataset and for less than 0.4\% of stars fainter than $G=10$. We therefore based our decision to use one proper motion catalogue or the other based on the cluster position only.


\begin{figure}[ht]
\begin{center} \resizebox{\hsize}{!}{\includegraphics[scale=0.5]{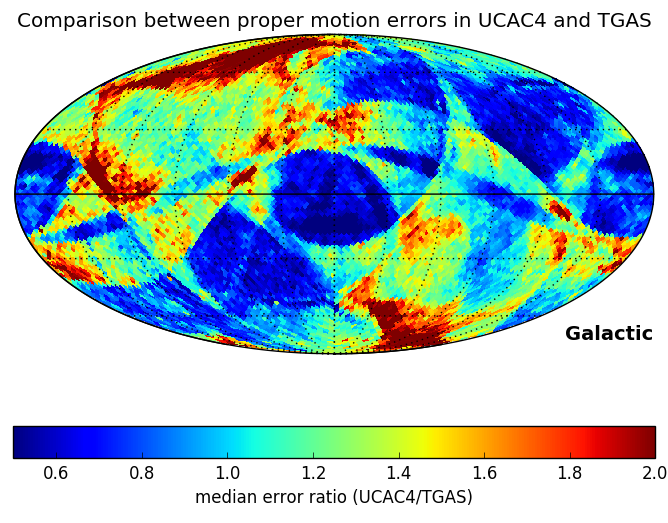}} \caption{\label{fig:mollweide_map_error_ratio} Median proper motion error ratio between UCAC4 and TGAS for stars present in both catalogues, in HEALPix level five pixels, in Galactic coordinates (north to the top, increasing longitude to the left, Galactic centre at the centre).} \end{center}
\end{figure}

\begin{figure}[ht]
\begin{center} \resizebox{\hsize}{!}{\includegraphics[scale=0.5]{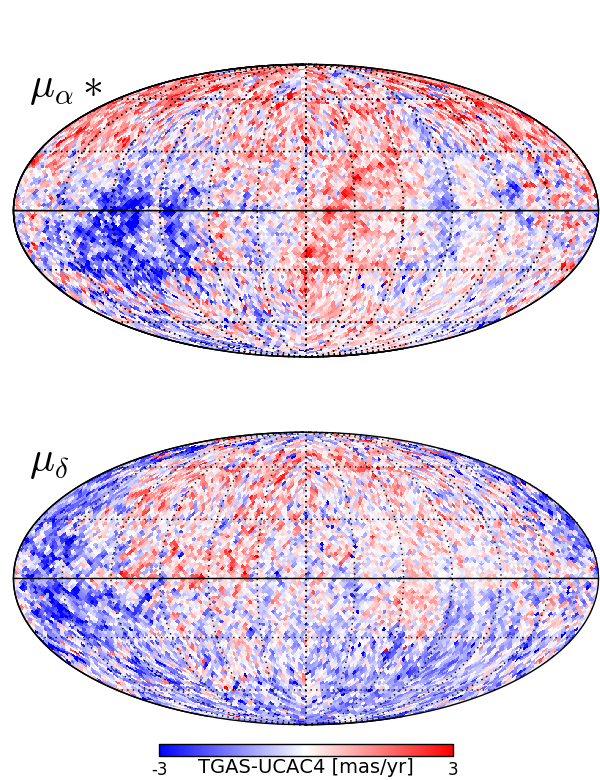}} \caption{\label{fig:map_PMdiff_TGAS_UCAC4} Top: Mean $\mu_{\alpha}*$ difference between UCAC4 and TGAS for stars present in both catalogues, in HEALPix level five pixels, in Galactic coordinates (north to the top, increasing longitude to the left, Galactic centre at the centre). Bottom: As for the $\mu_{\delta}$ component of the proper motion.} \end{center}
\end{figure}

\subsection{Target selection}
We obtained a list of cluster coordinates and parameters from the Milky Way Star Clusters catalogue of \citet[][hereafter MWSC]{Kharchenko13}.
Since in the TGAS data all stars with parallaxes under 0.5\,mas have relative errors $\sigma_{\varpi}/\varpi$ larger than 50\%, we limited our study to OCs with expected parallaxes larger than 0.5\,mas (closer than 2000\,pc). Since the TGAS stars are all contained in the Tycho-2 astrometric catalogue, it is affected by the same completeness limit of $V\sim11.5$. Making use of PARSEC stellar isochrones \citep{Bressan12}, we rejected all the clusters which, according to their listed age, distance, and extinction, were not expected to contain any star brighter than $V=12$.

Finally, excluded from this study the 19 nearby OCs studied in \citet{FvL17}, namely: the Hyades, the Pleiades, Coma Berenices, Praesepe, Alpha Perseus, Blanco~1, Collinder~140, IC~2391, IC~2602, IC~4665, NGC~2451A, NGC~6475, NGC~6633, NGC~7092, NGC~2516, NGC~2232, NGC~2422, NGC~3532, and NGC~2547.

This left us with a list of 694 OCs to investigate. For these objects, the MWSC quotes tidal radii of up to 15\,pc. To ensure we did not cut out potential cluster stars by applying selections that were too narrow, we kept all TGAS stars within an angle corresponding to a physical distance of 20\,pc at the reference distance of the cluster, and visually examined the sky distribution, parallax and proper motion distribution, and colour-magnitude for the members and field stars for each cluster in order to confirm the reality of the object.
When the procedure failed to identify any cluster member, we restricted the field of view to an angle corresponding to 10\,pc in order to provide a better contrast between field and cluster stars in the astrometric space. The final search radius used for each cluster is listed in Table~\ref{tab:meanparams}.

\subsection{Membership determination}

Our determination of cluster membership relies on proper motions and parallaxes.
To eliminate obvious field stars we first performed a broad selection, rejecting all stars with proper motions more distant than 10\,mas\,yr$^{-1}$ from the literature value.
For clusters closer than 500\,pc (expected parallax of 2\,mas) we also rejected stars with parallaxes under 1\,mas. About 90\% of the TGAS sources have parallax errors smaller than 0.5\,mas. From the parallax distribution error (shown in Fig.~\ref{fig:hist_par_eror}) we expect that only 1\% of TGAS sources have observed parallaxes that are lower than their true value by 1\,mas. This means that even for clusters with a true parallax of exactly 2\,mas, applying a cut-off at 1\,mas introduces no significant bias in the remaining parallax distribution.
Although the field of view of some pairs of OCs overlap (e.g. NGC~2451 with NGC~2477), 
 the difference in proper motions allowed us to tell the two clusters apart.

\begin{figure}[ht]
\begin{center} \resizebox{\hsize}{!}{\includegraphics[scale=0.5]{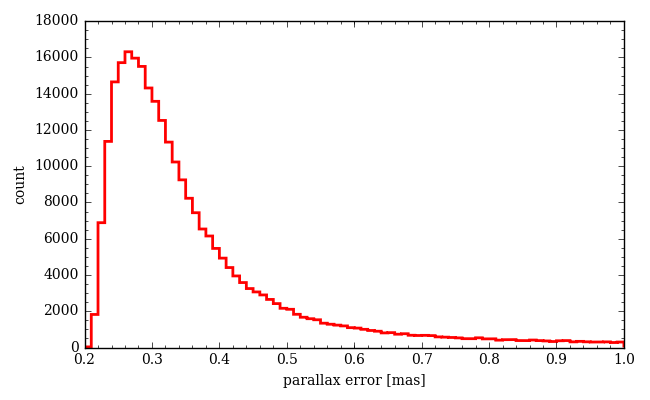}} \caption{\label{fig:hist_par_eror} Parallax error distribution (in bins of width 0.01\,mas) for the stars used in this study.} \end{center}
\end{figure}

For the remaining star sample, our approach to membership determination is based on the principles of the unsupervised membership assignment method UPMASK \citep[][]{KroneMartins14}. This approach does not rely on strong physical assumptions concerning the nature of a cluster (no assumptions on density profile modelling or on the structure in photometric space), except that its stars share common properties and that the spatial distributions of cluster stars is concentrated, while the distribution of field stars is random. In machine learning and data mining it is common to refer to a group of objects as a cluster. In this paper we tried to refer to the output of $k$-means clustering (abstract grouping of datapoints with similar observed properties) as groups, to avoid any possible confusion with stellar clusters (astronomical objects). The core idea of UPMASK is to apply a simple clustering algorithm (for instance, $k$-means clustering) to identify small groups of stars with similar colours and magnitudes in various photometric bands, then check all these small groups individually and determine whether their spatial distribution is more tightly concentrated than a random distribution (this is referred to as the ``veto'' step).

Although UPMASK was originally built to identify stellar clusters based on photometry \citep[and is used for this, for instance in][]{Costa15}, its core principle can easily be generalised to other types of quantities, as its only strong assumption is to consider that cluster members must be, in any observable space, more tightly distributed than field stars. This assumption happens to hold even truer for astrometry than photometry, as all stars within a cluster are expected to be located at the same distance from us and moving in the same direction, regardless of their colour and luminosity.
In this study, rather than applying the $k$-means clustering to a set of magnitudes in different photometric filters, we applied it to the three-dimensional astrometric space of proper motions and parallaxes $(\mu_{\alpha}*,\mu_{\delta},\varpi)$. As recommended in \citet{KroneMartins14}, we scale each of these three observables to unit variance. 
The $k$-means clustering method does not allow the user to impose the number of points in each group, as they can be of variable sizes, but it requires the user to choose the number of desired groups (which is equivalent to setting the mean number of stars in each group). \citet{KroneMartins14} report that they obtain the best results using values between ten and 25. In this study, since we expect several clusters to have very few members in the TGAS data, we set this mean number to ten.

To determine whether or not each identified group is spatially more concentrated than a random distribution, we applied the method introduced by \citet{Allison09} originally used to reveal mass segregation in star clusters (by showing whether high-mass stars are spatially more concentrated than the average of the cluster). 
The method consists in comparing the total branch length $ l_{obs} $ of the minimum spanning tree connecting all stars in that group to the expected value $l$ in a random distribution containing the same number of stars.

To save computation time, for sample sizes from three to 80 we pre-computed and tabulated the expected $l$ and associated standard deviations $\sigma_l$ by generating 2000 random circular distributions. If the $\langle l_{obs} \rangle $ value in a given group is smaller than the tabulated $l$ by at least $\sigma_l$, that is if

        \begin{equation} \label{eq:lambda}
                \Lambda = \frac{l - l_{obs} }{\sigma_l} > 1,
        \end{equation}

\noindent then all the stars in that group are flagged as possible cluster members. Figure~\ref{fig:ngc_0752_kmeans_veto} shows an example of $k$-means clustering applied to the stellar cluster NGC~752 and the sky distribution of three selected groups. In that example, two of the shown groups are considered as containing potential cluster members.

\begin{figure}[ht]
\begin{center} \resizebox{\hsize}{!}{\includegraphics[scale=0.5]{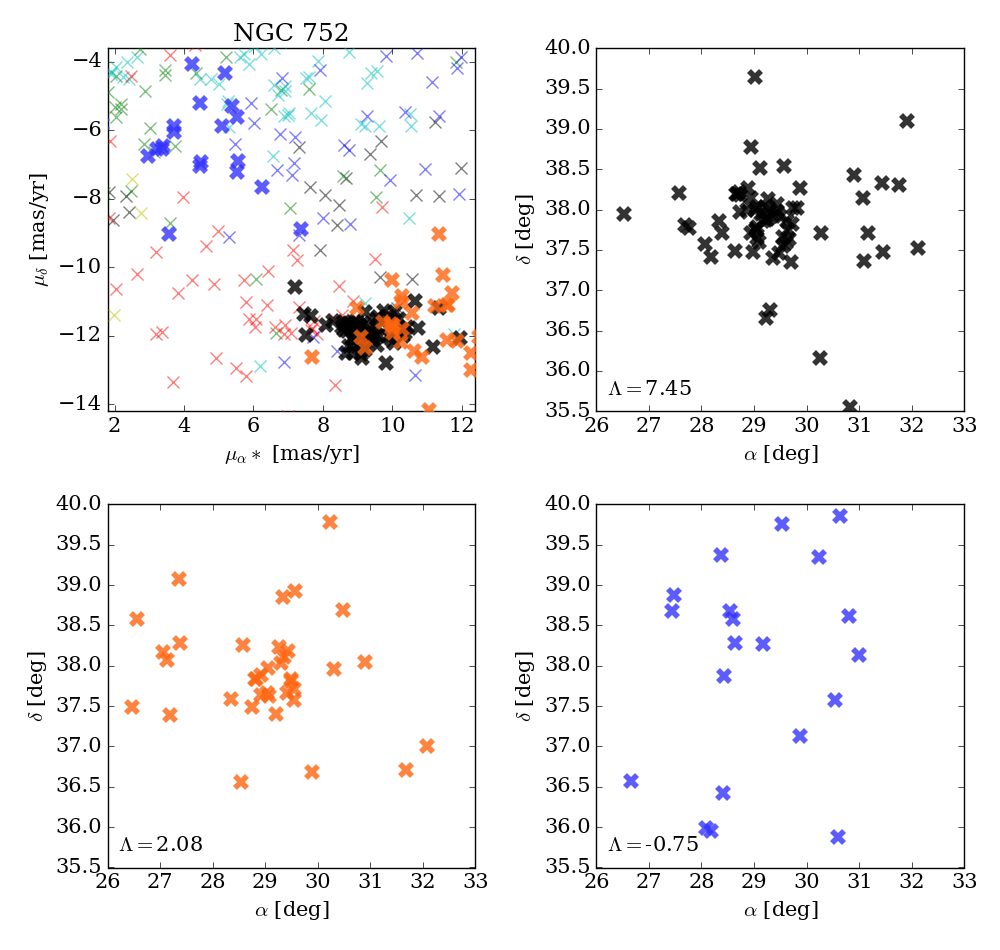}} \caption{\label{fig:ngc_0752_kmeans_veto} Top left panel: Proper motions of stars in field of NGC~752. The colour code corresponds to the groups identified by $k$-means clustering in the $(\mu_{\alpha}*,\mu_{\delta},\varpi)$ space. The sky distribution of the three highlighted groups is shown in the other panels. The quantity $\Lambda$ is a measurement of spatial clustering, as defined in Eq.~\ref{eq:lambda}. Stars are considered potential cluster members when $\Lambda>1$. } \end{center}
\end{figure}

When this veto step has been applied to all stars, the procedure ($k$-means clustering and spatial veto) is performed again, but instead of using the catalogue value we add to each datapoint a random offset in proper motion and parallax, corresponding to the  uncertainties. After a total of 100 iterations, the frequency with which a star was flagged as part of a clustered group is interpreted as its membership probability.

The random nature of the grouping step performed in the heuristic of UPMASK means that small groups of field stars might sometimes be flagged as clustered (in other words, even purely random distribution are expected to satisfy Eq.~\ref{eq:lambda}) and these stars end up with non-zero membership probabilities of a few percent, which can be considered noise level. To obtain cleaner results and a better contrast between field and cluster stars, we applied the procedure a second time for all OCs, after discarding stars for which the first run yielded probabilities lower than 10\%. We consider the final membership probability to be the result of this second run. The individual probabilities of all stars in the investigated fields of 128 clusters (including those with low or zero-probability) are provided as an electronic table. In total, we find 4450 potential cluster members (probability > 50\%) and 851 secure members (>90\%).
The result of the member selection procedure is shown in Fig.~\ref{fig:membersalessi3}  for the cluster Alessi~3 using TGAS proper motions and parallaxes.

\begin{figure}[ht]
\begin{center} \resizebox{\hsize}{!}{\includegraphics[scale=0.5]{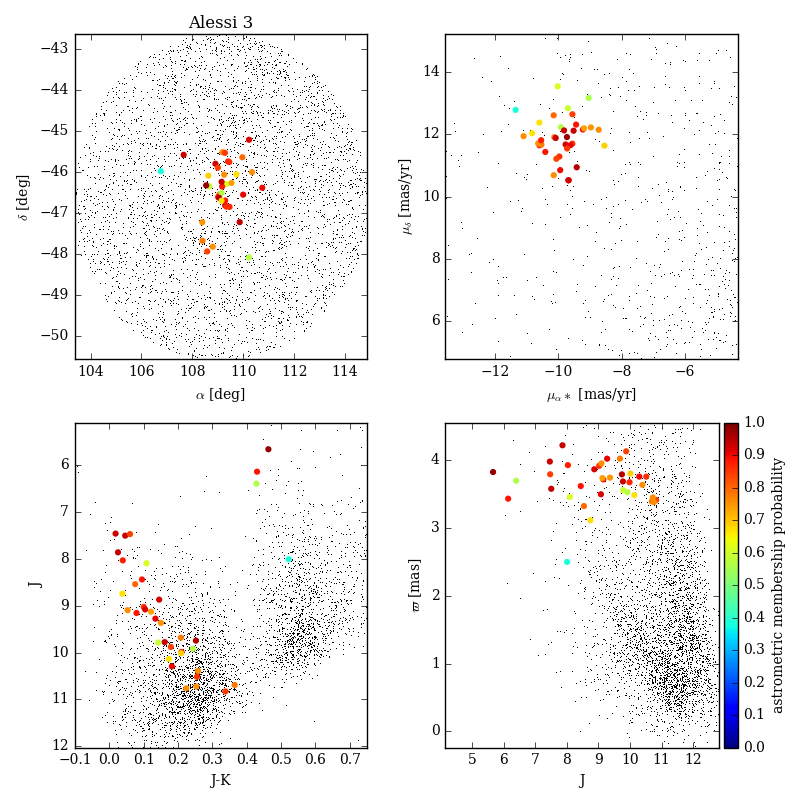}} \caption{\label{fig:membersalessi3} Top left: Positions of the stars identified as probable cluster members (dots coloured according to membership probability) and field stars (black points). Top right: Proper motions of the probable cluster members and of the field stars. Bottom left: $JK_S$ photometry for the cluster stars and field stars. Bottom right: Parallax $\varpi$ against $J$-magnitude for cluster and field stars. } \end{center}
\end{figure}

\subsection{Taking correlations into accounts}
Whenever possible (i.e. when the TGAS proper motions were used), the errors we added to the original values in each random drawing took into account the full covariance between the error in three parameters $(\mu_{\alpha}*,\mu_{\delta},\varpi)$. For each star, the covariance matrix is

        \begin{equation} \label{eq:covariancematrix}
        \mathrm{Cov} = 
                 \begin{bmatrix}
                        \sigma^{2}_{\mu_{\alpha}*}      & \sigma_{\mu_{\alpha}*} \sigma_{\mu_{\delta}} \rho_{\mu_{\alpha}*\mu_{\delta}}          & \sigma_{\mu_{\alpha}*} \sigma_{\varpi} \rho_{\mu_{\alpha}*\varpi}      \\
                        \sigma_{\mu_{\alpha}*} \sigma_{\mu_{\delta}} \rho_{\mu_{\alpha}*\mu_{\delta}}   &  \sigma^{2}_{\mu_{\delta}}      &  \sigma_{\mu_{\delta}} \sigma_{\varpi} \rho_{ \mu_{\delta} \varpi}     \\
                        \sigma_{\mu_{\alpha}*} \sigma_{\varpi} \rho_{\mu_{\alpha}*\varpi}       &  \sigma_{\mu_{\delta}} \sigma_{\varpi} \rho_{ \mu_{\delta} \varpi}      & \sigma^{2}_{\varpi}     \\
                 \end{bmatrix} ,
        \end{equation}

\noindent where $\sigma_{\mu_{\alpha}*}$, $\sigma_{\mu_{\delta}}$, and $\sigma_{\varpi}$,
and $\rho_{\mu_{\alpha}*\mu_{\delta}}$, $\rho_{\mu_{\alpha}*\varpi}$, and $\rho_{ \mu_{\delta} \varpi}$ (the correlation coefficients) are all provided in the TGAS catalogue. 
Neglecting the off-diagonal terms is equivalent to considering the correlation coefficients $\rho$ as equal to zero (i.e. that the errors are independent), which is generally not true.  This is illustrated in Fig.~\ref{fig:histograms_rho}, where $\rho$ for proper motions and parallaxes are shown to be significantly different from zero for a large number of stars used in this study. Indeed, the correlations between the errors of astrometric parameters are expected to vary in different regions of the sky, as explained in \citet{Lindegren16} (see for instance   the map in their Fig.~7).

\begin{figure*}[ht]
\begin{center} \resizebox{\hsize}{!}{\includegraphics[scale=0.5]{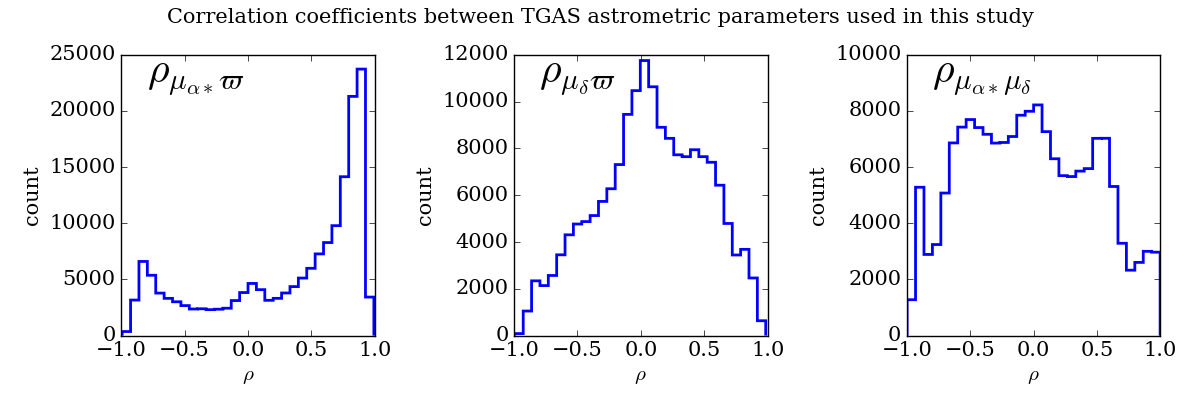}} \caption{\label{fig:histograms_rho} Histograms for the value of the correlation coefficient $\rho$ between the three astrometric parameters $\mu_{\alpha}*$,$\mu_{\delta}$, and $\varpi$, for the stars used in this study in the fields where the TGAS proper motions uncertainties are smaller than the UCAC4 uncertainties. } \end{center}
\end{figure*}

A visual example is shown in Fig.~\ref{fig:ngc2360_pmcorrelations}, where the bottom left and bottom middle panels show the proper motions of stars in NGC~2360, displaying the uncertainties as if $\mu_{\alpha}*$ and $\mu_{\delta}$ errors were uncorrelated. The representation in the bottom right of Fig.~\ref{fig:ngc2360_pmcorrelations} takes into consideration the correlations between both components of the proper motion, showing the uncertainties as covariant, tilted ellipses, which when correlations are non-zero are always narrower than the non-covariant representation. As a result, some stars that appear as marginally compatible with being cluster members in the bottom left panel of Fig.~\ref{fig:ngc2360_pmcorrelations} clearly appear as outliers in the bottom right panel.

When using UCAC4 proper motions, for which the correlations between the errors on the components of the proper motion are not available, we set the correlations coefficients to zero. We also assumed that the error on the TGAS parallax is uncorrelated with the UCAC4 proper motion errors for these stars.

\begin{figure*}[ht]
\begin{center} \resizebox{\hsize}{!}{\includegraphics[scale=0.5]{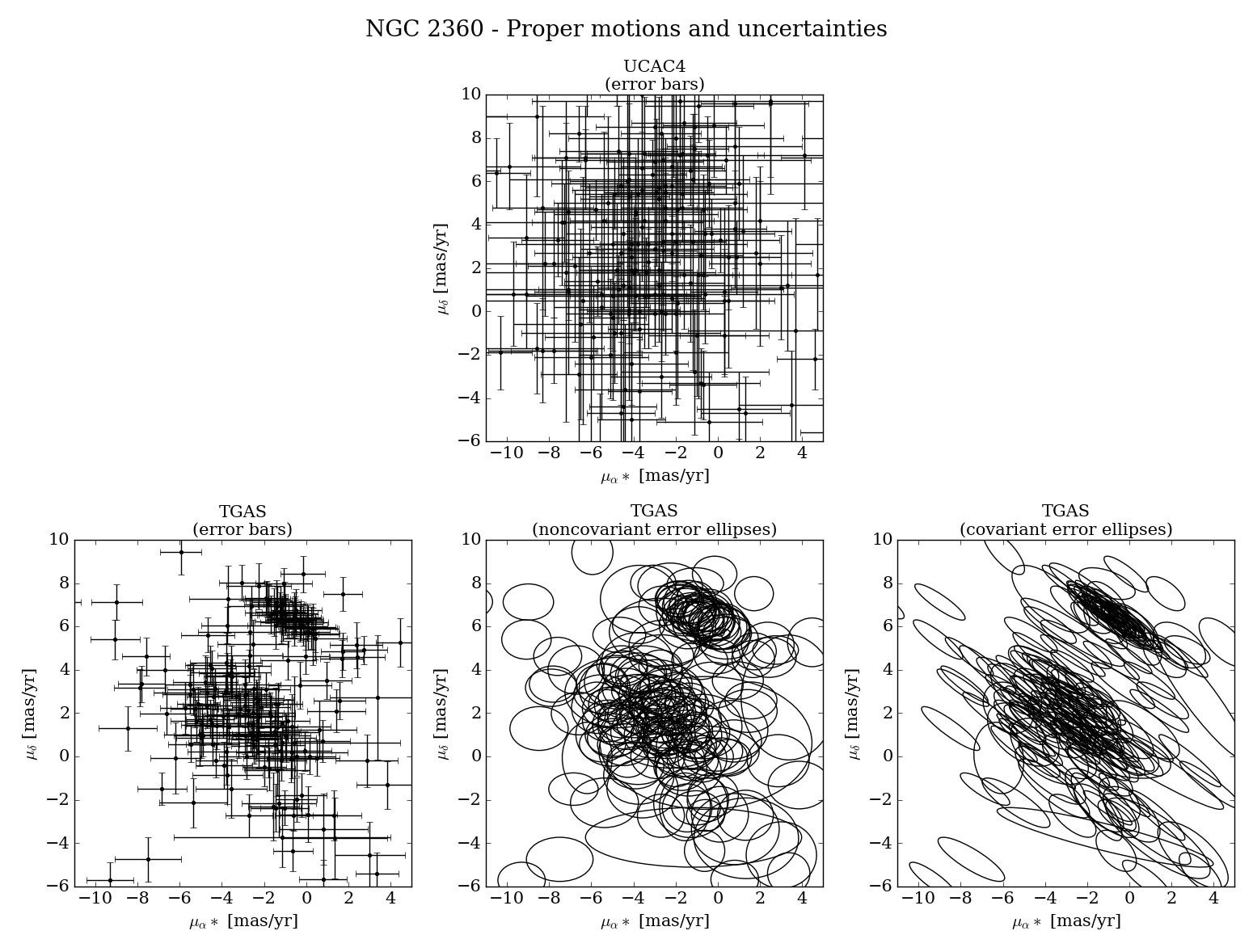}} \caption{\label{fig:ngc2360_pmcorrelations} Top: UCAC4 proper motions for NGC\,2360. Bottom left: TGAS proper motions for the same set of stars, with nominal uncertainties represented as error bars. Bottom middle: Same as bottom left panel, but the uncertainties are represented as non-covariant error ellipses. Bottom right: Same data as bottom middle panel, with uncertainties represented as tilted error ellipses, showing the correlations between $\mu_{\alpha}*$ and $\mu_{\delta}$ errors (here of the order of $-0.8$). } \end{center}
\end{figure*}

\section{Mean proper motions and parallaxes} \label{sec:meanparams}

\subsection{Computation}
 
We computed the mean astrometric parameters of the cluster members after applying a 2-$\sigma$ clipping on the proper motions, parallax, and sky distribution of the member stars. 
In order to determine a mean proper motion and mean parallax for each cluster taking into account both the membership probability and the uncertainty on the parameters of each star, we performed a thousand random drawings where we picked stars according to their probability of being cluster members. 
For each redrawing, the mean proper motion and mean parallax was computed as a weighted mean, where the weight for each star corresponds to the inverse variance (inverse  uncertainty, squared) and the uncertainty on this mean value is the standard deviation divided by the square root of the number of selected stars. The final value is the average over all redrawings, and the final uncertainty is the quadratic sum of the average the uncertainty and the standard deviation of the mean value over all redrawings. 

In this computation of mean parameters we do not take into account the correlations between astrometric parameters for individual stars, as they do not have a significant impact on the mean values and their uncertainties. However, due to the fact that correlations tend to be similar among sources in a given field and therefore do not cancel out, the mean parameters themselves are correlated, and a more complete treatment should be applied if one was interested in determining the full covariance of the mean parameters. As a result, the volume of parameter space covered by the possible combinations of $\langle \varpi \rangle$, $\langle \mu_{\alpha}* \rangle,$ and $\langle \mu_{\delta} \rangle$ is smaller when correlations exist between these quantities.

The final results are listed in Tables~\ref{tab:meanparams} and~\ref{tab:meanparams2}. Since the number of identified members is rather low for many clusters, we did not update the coordinates of the cluster centre. In all cases, the positions of the member stars were consistent with the literature data.

\begin{table*}
\begin{center}
        \caption{ \label{tab:meanparams} Mean astrometric parameters computed for 128 OCs.}
        \small\addtolength{\tabcolsep}{-1pt}
        \begin{tabular}{l l l l l l l l l l l l l l l l l l}
        \hline
        \hline
   OC   & $\alpha$ & $\delta$ & $\langle \varpi \rangle$      & $\varpi_{\textrm{DAML}}$ & $\varpi_{\textrm{MWSC}}$ & $\langle \mu_{\alpha}* \rangle$  &  $\langle \mu_{\delta} \rangle$  & PM$\dagger$  & Search radius & nb\\
        & [deg]    & [deg]    & [mas]                         & [mas]               & [mas]               & [mas\,yr$^{-1}$]                 &  [mas\,yr$^{-1}$]                &              &  [arcmin]     &   \\
        \hline
        \object{ASCC~10} & 51.765 & 35.030 & 1.92 $\pm$ 0.20 & -- & 1.43 & -3.38 $\pm$ 0.16 & -1.13 $\pm$ 0.20 & U & 97 & 21 \\
        \object{ASCC~112} & 304.095 & 52.098 & 1.52 $\pm$ 0.27 & -- & 1.76 & -0.96 $\pm$ 0.31 & -0.45 $\pm$ 0.33 & U & 61 & 4 \\
        \object{ASCC~113} & 318.007 & 38.610 & 1.74 $\pm$ 0.07 & 2.22 & 2.0 & 0.78 $\pm$ 0.13 & -3.86 $\pm$ 0.09 & T & 74 & 35 \\
        \object{ASCC~123} & 340.597 & 54.250 & 4.58 $\pm$ 0.16 & 4.0 & 4.0 & 11.42 $\pm$ 0.28 & -1.54 $\pm$ 0.29 & U & 136 & 9 \\
        \object{ASCC~124} & 341.895 & 46.310 & 1.27 $\pm$ 0.13 & -- & 1.43 & 0.36 $\pm$ 0.10 & -1.85 $\pm$ 0.10 & T & 54 & 11 \\
        \object{ASCC~16} & 81.037 & 1.640 & 2.89 $\pm$ 0.12 & -- & 2.52 & 0.92 $\pm$ 0.28 & 0.56 $\pm$ 0.21 & U & 70 & 27 \\
        \object{ASCC~18} & 81.562 & 0.900 & 2.89 $\pm$ 0.10 & 2.0 & 3.19 & 0.54 $\pm$ 0.24 & 0.50 $\pm$ 0.14 & U & 133 & 45 \\
        \object{ASCC~19} & 81.922 & -1.965 & 2.69 $\pm$ 0.12 & 2.86 & 3.18 & 0.97 $\pm$ 0.18 & -0.97 $\pm$ 0.15 & T & 96 & 25 \\
        \object{ASCC~21} & 82.260 & 3.560 & 3.11 $\pm$ 0.15 & 2.0 & 2.64 & 0.60 $\pm$ 0.23 & 0.79 $\pm$ 0.28 & U & 66 & 13 \\
        \object{ASCC~23} & 95.089 & 46.675 & 1.59 $\pm$ 0.14 & 1.67 & 1.67 & 0.23 $\pm$ 0.24 & -0.48 $\pm$ 0.11 & U & 112 & 18 \\
        \object{ASCC~32} & 105.525 & -26.560 & 1.23 $\pm$ 0.08 & -- & 1.3 & -3.36 $\pm$ 0.10 & 3.68 $\pm$ 0.08 & T & 89 & 33 \\
        \object{ASCC~41} & 116.745 & 0.020 & 3.30 $\pm$ 0.17 & -- & 2.52 & 1.13 $\pm$ 0.18 & -4.45 $\pm$ 0.63 & U & 89 & 12 \\
        \object{ASCC~51} & 139.522 & -69.690 & 1.99 $\pm$ 0.29 & 2.0 & 2.02 & -8.12 $\pm$ 0.35 & 8.30 $\pm$ 0.29 & T & 66 & 6 \\
        \object{ASCC~99} & 282.277 & -18.740 & 3.37 $\pm$ 0.15 & 3.57 & 1.47 & 5.72 $\pm$ 0.48 & -0.49 $\pm$ 0.33 & U & 119 & 15 \\
        \object{Alessi~10} & 301.182 & -10.485 & 1.92 $\pm$ 0.19 & 1.95 & 1.95 & 1.90 $\pm$ 0.28 & -7.10 $\pm$ 0.37 & T & 63 & 11 \\
        \object{Alessi~12} & 310.845 & 23.785 & 1.90 $\pm$ 0.13 & 1.86 & 1.84 & 3.66 $\pm$ 0.20 & -5.21 $\pm$ 0.15 & U & 126 & 21 \\
        \object{Alessi~13} & 52.020 & -35.870 & 9.50 $\pm$ 0.44 & 10.0 & 8.93 & 36.43 $\pm$ 0.56 & -4.41 $\pm$ 0.33 & T & 624 & 15 \\
        \object{Alessi~21} & 107.670 & -9.300 & 1.72 $\pm$ 0.09 & 2.0 & 1.68 & -5.41 $\pm$ 0.19 & 2.53 $\pm$ 0.20 & T & 66 & 31 \\
        \object{Alessi~2} & 71.565 & 55.255 & 1.69 $\pm$ 0.09 & 2.0 & 1.92 & -0.86 $\pm$ 0.27 & -0.80 $\pm$ 0.18 & U & 62 & 21 \\
        \object{Alessi~3} & 109.050 & -46.617 & 3.71 $\pm$ 0.08 & 3.47 & 3.22 & -9.70 $\pm$ 0.11 & 12.21 $\pm$ 0.12 & T & 237 & 29 \\
        \object{Alessi~5} & 160.830 & -61.170 & 2.51 $\pm$ 0.04 & 2.51 & 2.6 & -15.50 $\pm$ 0.08 & 2.62 $\pm$ 0.11 & T & 84 & 16 \\
        \object{Alessi~6} & 220.132 & -66.120 & 1.23 $\pm$ 0.03 & -- & 1.63 & -10.36 $\pm$ 0.09 & -5.67 $\pm$ 0.08 & T & 76 & 21 \\
        \object{Alessi~9} & 266.295 & -47.185 & 4.89 $\pm$ 0.13 & 4.74 & 4.59 & 11.10 $\pm$ 0.33 & -8.21 $\pm$ 0.41 & U & 351 & 25 \\
        \object{BH~99} & 159.465 & -59.125 & 2.39 $\pm$ 0.15 & 1.97 & 1.93 & -14.48 $\pm$ 0.12 & 1.30 $\pm$ 0.13 & T & 67 & 12 \\
        \object{Chereul~1} & 217.267 & 55.392 & 10.15 $\pm$ 0.31 & -- & -- & -17.89 $\pm$ 0.32 & -5.01 $\pm$ 0.38 & T & 653 & 8 \\
        \object{Collinder~135} & 109.485 & -37.035 & 3.46 $\pm$ 0.12 & 3.16 & 2.77 & -9.28 $\pm$ 0.27 & 5.57 $\pm$ 0.22 & U & 217 & 36 \\
        \object{Collinder~350} & 267.052 & 1.360 & 2.71 $\pm$ 0.13 & 3.57 & 3.31 & -3.68 $\pm$ 0.35 & 0.29 $\pm$ 0.29 & U & 242 & 19 \\
        \object{Collinder~359} & 270.120 & 2.875 & 1.93 $\pm$ 0.10 & 4.02 & 1.56 & 1.98 $\pm$ 0.23 & -8.19 $\pm$ 0.26 & U & 271 & 54 \\
        \object{Collinder~394} & 283.080 & -20.200 & 1.38 $\pm$ 0.10 & 1.45 & 1.5 & -0.76 $\pm$ 0.28 & -4.81 $\pm$ 0.32 & U & 45 & 32 \\
        \object{Collinder~463} & 27.300 & 71.780 & 1.36 $\pm$ 0.07 & 1.42 & 1.25 & -1.83 $\pm$ 0.07 & -0.29 $\pm$ 0.15 & T & 96 & 41 \\
        \object{IC~4725} & 277.927 & -19.110 & 1.52 $\pm$ 0.12 & 1.61 & 1.78 & -3.43 $\pm$ 0.35 & -5.87 $\pm$ 0.33 & U & 50 & 44 \\
        \object{IC~4756} & 279.735 & 5.450 & 1.94 $\pm$ 0.03 & 2.07 & 2.07 & 0.66 $\pm$ 0.13 & -2.58 $\pm$ 0.14 & U & 141 & 99 \\
        \object{Lynga~2} & 216.112 & -61.330 & 0.97 $\pm$ 0.12 & 1.11 & 1.07 & -6.76 $\pm$ 0.19 & -4.65 $\pm$ 0.23 & T & 32 & 9 \\
        \object{Melotte~101} & 160.567 & -65.095 & 0.41 $\pm$ 0.18 & 0.5 & 0.5 & -6.91 $\pm$ 0.24 & 2.47 $\pm$ 0.18 & T & 31 & 10 \\
        \object{NGC~0752} & 29.272 & 37.790 & 2.34 $\pm$ 0.05 & -- & 2.22 & 8.19 $\pm$ 0.09 & -12.12 $\pm$ 0.07 & U & 148 & 73 \\
        \object{NGC~1027} & 40.665 & 61.620 & 0.74 $\pm$ 0.09 & 0.97 & 1.05 & -2.91 $\pm$ 0.22 & 2.11 $\pm$ 0.16 & U & 42 & 21 \\
        \object{NGC~1039} & 40.549 & 42.790 & 1.94 $\pm$ 0.10 & 2.0 & 1.96 & -0.51 $\pm$ 0.17 & -6.39 $\pm$ 0.13 & U & 134 & 41 \\
        \object{NGC~1342} & 53.004 & 37.357 & 1.47 $\pm$ 0.06 & 1.5 & 1.5 & 0.08 $\pm$ 0.19 & -1.54 $\pm$ 0.08 & T & 49 & 22 \\
        \object{NGC~1528} & 63.847 & 51.190 & 0.99 $\pm$ 0.09 & 0.92 & 1.05 & 2.08 $\pm$ 0.19 & -1.71 $\pm$ 0.16 & U & 40 & 21 \\
        \object{NGC~1545} & 65.295 & 50.252 & 1.38 $\pm$ 0.14 & 1.41 & 1.16 & -2.26 $\pm$ 0.17 & 0.45 $\pm$ 0.18 & U & 45 & 10 \\
        \object{NGC~1647} & 71.497 & 19.170 & 1.88 $\pm$ 0.04 & 1.85 & 1.75 & -0.97 $\pm$ 0.09 & -1.66 $\pm$ 0.12 & U & 122 & 57 \\
        \object{NGC~1662} & 72.112 & 10.925 & 2.56 $\pm$ 0.07 & 2.29 & 2.29 & -1.05 $\pm$ 0.14 & -2.02 $\pm$ 0.11 & T & 154 & 36 \\
        \object{NGC~1750} & 76.140 & 23.830 & 1.48 $\pm$ 0.08 & 1.59 & -- & -3.39 $\pm$ 0.14 & -3.71 $\pm$ 0.13 & U & 107 & 57 \\
        \object{NGC~1778} & 76.995 & 37.028 & 0.52 $\pm$ 0.11 & 0.68 & 0.67 & -0.37 $\pm$ 0.41 & -4.42 $\pm$ 0.38 & U & 18 & 5 \\
        \object{NGC~1901} & 79.560 & -68.440 & 2.51 $\pm$ 0.11 & 2.17 & 2.46 & 0.78 $\pm$ 0.31 & 13.11 $\pm$ 0.24 & T & 81 & 6 \\
        \object{NGC~1912} & 82.215 & 35.800 & 0.88 $\pm$ 0.07 & 0.71 & 0.87 & -0.01 $\pm$ 0.23 & -4.51 $\pm$ 0.25 & U & 29 & 39 \\
        \object{NGC~1960} & 84.082 & 34.158 & 1.05 $\pm$ 0.07 & 0.75 & 0.83 & -0.50 $\pm$ 0.17 & -5.29 $\pm$ 0.14 & U & 21 & 26 \\
        \object{NGC~1977} & 83.887 & -4.830 & 2.31 $\pm$ 0.15 & 2.0 & 2.79 & -0.25 $\pm$ 0.23 & -0.17 $\pm$ 0.32 & T & 64 & 14 \\
        \object{NGC~2099} & 88.087 & 32.570 & 0.76 $\pm$ 0.11 & 0.72 & 0.71 & 2.25 $\pm$ 0.24 & -7.12 $\pm$ 0.20 & U & 48 & 30 \\
        \object{NGC~2168} & 92.302 & 24.360 & 1.12 $\pm$ 0.05 & 1.1 & 1.2 & 0.62 $\pm$ 0.10 & -4.06 $\pm$ 0.08 & U & 83 & 103 \\
        \object{NGC~2215} & 95.220 & -7.295 & 1.31 $\pm$ 0.26 & 1.27 & 0.88 & 0.19 $\pm$ 0.29 & -3.12 $\pm$ 0.34 & T & 52 & 16 \\
        \object{NGC~2244} & 97.980 & 4.940 & 0.86 $\pm$ 0.13 & 0.6 & 0.65 & -0.68 $\pm$ 0.51 & 0.56 $\pm$ 0.52 & U & 20 & 8 \\
        \object{NGC~2264} & 100.245 & 9.880 & 1.38 $\pm$ 0.11 & 1.5 & 1.54 & -0.51 $\pm$ 0.30 & -3.68 $\pm$ 0.28 & U & 48 & 6 \\
        \object{NGC~2281} & 102.075 & 41.080 & 2.03 $\pm$ 0.08 & 1.79 & 2.0 & -3.96 $\pm$ 0.14 & -8.05 $\pm$ 0.14 & U & 121 & 35 \\
        \object{NGC~2287} & 101.512 & -20.745 & 1.29 $\pm$ 0.03 & 1.41 & 1.3 & -4.40 $\pm$ 0.06 & -1.39 $\pm$ 0.04 & T & 97 & 77 \\
        \object{NGC~2323} & 105.679 & -8.370 & 1.01 $\pm$ 0.04 & 1.05 & 1.11 & -0.69 $\pm$ 0.11 & -0.35 $\pm$ 0.10 & T & 72 & 110 \\
        \object{NGC~2353} & 108.627 & -10.250 & 0.67 $\pm$ 0.16 & 0.85 & 0.85 & -1.03 $\pm$ 0.25 & 0.81 $\pm$ 0.24 & T & 29 & 18 \\
        \object{NGC~2360} & 109.425 & -15.640 & 0.77 $\pm$ 0.04 & 0.89 & 0.64 & -0.14 $\pm$ 0.16 & 5.99 $\pm$ 0.11 & T & 35 & 40 \\
        \object{NGC~2423} & 114.277 & -13.850 & 1.07 $\pm$ 0.06 & 1.31 & 1.21 & -0.82 $\pm$ 0.11 & -3.54 $\pm$ 0.09 & T & 78 & 78 \\
        \object{NGC~2437} & 115.455 & -14.805 & 0.66 $\pm$ 0.03 & 0.66 & 0.73 & -3.78 $\pm$ 0.08 & 0.33 $\pm$ 0.06 & T & 49 & 120 \\
        \object{NGC~2447} & 116.145 & -23.860 & 1.13 $\pm$ 0.06 & 0.96 & 0.96 & -3.51 $\pm$ 0.15 & 5.06 $\pm$ 0.06 & T & 65 & 62 \\
        \object{NGC~2451B} & 116.115 & -37.670 & 2.64 $\pm$ 0.10 & 3.31 & 2.04 & -9.70 $\pm$ 0.26 & 4.33 $\pm$ 0.13 & T & 227 & 34 \\
        \object{NGC~2477} & 118.035 & -38.530 & 0.64 $\pm$ 0.12 & 0.75 & 0.69 & -0.70 $\pm$ 0.57 & 1.94 $\pm$ 0.52 & U & 25 & 18 \\
        \object{NGC~2482} & 118.792 & -24.275 & 0.75 $\pm$ 0.11 & 0.84 & 0.74 & -4.47 $\pm$ 0.24 & 2.27 $\pm$ 0.12 & T & 23 & 35 \\
        \object{NGC~2527} & 121.275 & -28.170 & 1.71 $\pm$ 0.04 & 1.66 & 1.56 & -5.38 $\pm$ 0.08 & 7.37 $\pm$ 0.08 & T & 113 & 83 \\
        \hline
        \end{tabular}
\tablefoot{Flag $\dagger$U indicates that the UCAC4 proper motions were used, T indicates TGAS proper motions. $ \varpi_{\textrm{DAML}}$ and $\varpi_{\textrm{MWSC}}$ are the expected parallaxes assuming no error on the catalogue value.}
\end{center}
\end{table*}

\begin{table*}
\begin{center}
        \caption{ \label{tab:meanparams2} Table~\ref{tab:meanparams}, continued.}
        \small\addtolength{\tabcolsep}{-1pt}
        \begin{tabular}{l l l l l l l l l l l l l l l l l l}
        \hline
        \hline
   OC   & $\alpha$ & $\delta$ & $\langle \varpi \rangle$      & $\varpi_{\textrm{DAML}}$ & $\varpi_{\textrm{MWSC}}$ & $\langle \mu_{\alpha}* \rangle$  &  $\langle \mu_{\delta} \rangle$  & PM$\dagger$  & Search radius & nb\\
        & [deg]    & [deg]    & [mas]                         & [mas]               & [mas]               & [mas\,yr$^{-1}$]                 &  [mas\,yr$^{-1}$]                &              &  [arcmin]     &   \\
        \hline
        \object{NGC~2539} & 122.670 & -12.840 & 1.17 $\pm$ 0.18 & 0.73 & 0.81 & -2.48 $\pm$ 0.25 & -2.33 $\pm$ 0.29 & U & 47 & 25 \\
        \object{NGC~2546} & 122.962 & -37.610 & 0.98 $\pm$ 0.07 & 1.09 & 1.07 & -4.30 $\pm$ 0.11 & 4.41 $\pm$ 0.10 & T & 35 & 46 \\
        \object{NGC~2548} & 123.435 & -5.770 & 1.44 $\pm$ 0.06 & 1.3 & 1.27 & -0.39 $\pm$ 0.20 & 2.68 $\pm$ 0.23 & U & 87 & 63 \\
        \object{NGC~2567} & 124.642 & -30.650 & 0.56 $\pm$ 0.06 & 0.6 & 0.6 & -2.97 $\pm$ 0.11 & 2.34 $\pm$ 0.08 & T & 19 & 21 \\
        \object{NGC~2571} & 124.735 & -29.735 & 1.02 $\pm$ 0.14 & 0.75 & 0.75 & -4.50 $\pm$ 0.21 & 4.36 $\pm$ 0.14 & T & 23 & 14 \\
        \object{NGC~2669} & 131.617 & -52.940 & 1.06 $\pm$ 0.16 & 0.96 & 0.95 & -4.09 $\pm$ 0.35 & 4.55 $\pm$ 0.40 & T & 30 & 11 \\
        \object{NGC~2670} & 131.392 & -48.820 & 0.59 $\pm$ 0.09 & 0.84 & 0.77 & -5.68 $\pm$ 0.20 & 3.70 $\pm$ 0.23 & T & 24 & 14 \\
        \object{NGC~2682} & 132.847 & 11.815 & 1.42 $\pm$ 0.20 & 1.24 & 1.12 & -9.38 $\pm$ 0.35 & -4.35 $\pm$ 0.33 & U & 31 & 17 \\
        \object{NGC~3228} & 155.355 & -51.730 & 2.27 $\pm$ 0.11 & 1.84 & 1.75 & -15.63 $\pm$ 0.48 & -0.34 $\pm$ 0.38 & T & 125 & 19 \\
        \object{NGC~3330} & 159.690 & -54.115 & 0.65 $\pm$ 0.07 & 1.12 & 0.8 & -7.20 $\pm$ 0.13 & 0.97 $\pm$ 0.14 & T & 36 & 21 \\
        \object{NGC~3680} & 171.405 & -43.235 & 1.18 $\pm$ 0.10 & 1.07 & 1.06 & -7.15 $\pm$ 0.22 & 0.90 $\pm$ 0.18 & T & 34 & 17 \\
        \object{NGC~4103} & 181.665 & -61.250 & 0.38 $\pm$ 0.27 & 0.61 & 0.6 & -6.56 $\pm$ 0.31 & -0.60 $\pm$ 0.24 & T & 17 & 6 \\
        \object{NGC~4609} & 190.575 & -62.990 & 0.60 $\pm$ 0.17 & 0.76 & 0.76 & -4.97 $\pm$ 0.27 & -1.20 $\pm$ 0.19 & T & 25 & 8 \\
        \object{NGC~4852} & 195.045 & -59.590 & 0.63 $\pm$ 0.14 & 0.91 & 0.9 & -8.03 $\pm$ 0.14 & -2.10 $\pm$ 0.19 & T & 29 & 21 \\
        \object{NGC~5138} & 201.795 & -59.030 & 0.62 $\pm$ 0.11 & 0.5 & 0.55 & -3.67 $\pm$ 0.10 & -1.41 $\pm$ 0.20 & T & 16 & 7 \\
        \object{NGC~5316} & 208.485 & -61.870 & 0.67 $\pm$ 0.05 & 0.82 & 0.83 & -6.34 $\pm$ 0.07 & -1.53 $\pm$ 0.07 & T & 25 & 29 \\
        \object{NGC~5460} & 211.890 & -48.330 & 1.49 $\pm$ 0.18 & 1.43 & 1.53 & -3.74 $\pm$ 0.50 & -0.51 $\pm$ 0.30 & U & 48 & 19 \\
        \object{NGC~5617} & 217.447 & -60.715 & 0.38 $\pm$ 0.08 & 0.5 & 0.57 & -5.38 $\pm$ 0.17 & -3.20 $\pm$ 0.08 & T & 21 & 24 \\
        \object{NGC~5662} & 218.895 & -56.615 & 1.37 $\pm$ 0.06 & 1.5 & 1.6 & -6.54 $\pm$ 0.14 & -7.18 $\pm$ 0.07 & T & 49 & 31 \\
        \object{NGC~5822} & 226.117 & -54.390 & 1.19 $\pm$ 0.04 & 1.07 & 1.26 & -7.44 $\pm$ 0.09 & -5.30 $\pm$ 0.05 & T & 74 & 96 \\
        \object{NGC~6025} & 240.817 & -60.400 & 1.21 $\pm$ 0.05 & 1.32 & 1.31 & -3.05 $\pm$ 0.12 & -3.00 $\pm$ 0.10 & T & 44 & 35 \\
        \object{NGC~6067} & 243.300 & -54.220 & 0.37 $\pm$ 0.06 & 0.71 & 0.56 & -1.98 $\pm$ 0.15 & -2.78 $\pm$ 0.08 & T & 22 & 46 \\
        \object{NGC~6087} & 244.710 & -57.940 & 0.93 $\pm$ 0.05 & 1.12 & 1.12 & -1.67 $\pm$ 0.08 & -2.47 $\pm$ 0.08 & T & 36 & 15 \\
        \object{NGC~6124} & 246.315 & -40.650 & 1.66 $\pm$ 0.05 & 1.95 & 1.79 & -0.53 $\pm$ 0.20 & -0.53 $\pm$ 0.21 & U & 130 & 48 \\
        \object{NGC~6134} & 246.952 & -49.155 & 1.03 $\pm$ 0.11 & 0.79 & 1.12 & 0.77 $\pm$ 0.38 & -4.18 $\pm$ 0.41 & T & 36 & 20 \\
        \object{NGC~6152} & 248.175 & -52.640 & 0.63 $\pm$ 0.05 & 0.97 & 0.97 & -3.14 $\pm$ 0.20 & -4.88 $\pm$ 0.20 & T & 30 & 30 \\
        \object{NGC~6281} & 256.170 & -37.980 & 1.70 $\pm$ 0.08 & 2.09 & 1.95 & -2.04 $\pm$ 0.26 & -2.81 $\pm$ 0.32 & U & 70 & 57 \\
        \object{NGC~6405} & 265.080 & -32.215 & 2.00 $\pm$ 0.14 & 2.05 & 2.81 & -1.44 $\pm$ 0.31 & -4.76 $\pm$ 0.31 & U & 68 & 41 \\
        \object{NGC~6416} & 266.092 & -32.365 & 1.06 $\pm$ 0.12 & 1.35 & 1.3 & -1.09 $\pm$ 0.46 & -0.47 $\pm$ 0.36 & U & 43 & 28 \\
        \object{NGC~6494} & 269.227 & -19.000 & 1.23 $\pm$ 0.08 & 1.59 & 1.54 & 1.17 $\pm$ 0.24 & 0.09 $\pm$ 0.22 & U & 49 & 53 \\
        \object{NGC~6604} & 274.507 & -12.238 & 0.54 $\pm$ 0.15 & 0.59 & 0.53 & -0.72 $\pm$ 0.36 & -0.97 $\pm$ 0.46 & U & 37 & 23 \\
        \object{NGC~6694} & 281.317 & -9.380 & 0.51 $\pm$ 0.05 & 0.6 & 0.57 & 0.28 $\pm$ 0.28 & -1.01 $\pm$ 0.24 & T & 17 & 12 \\
        \object{NGC~6705} & 282.747 & -6.280 & 0.57 $\pm$ 0.13 & 0.53 & 0.57 & -2.19 $\pm$ 0.46 & -5.07 $\pm$ 0.48 & T & 15 & 7 \\
        \object{NGC~6716} & 283.642 & -19.885 & 1.38 $\pm$ 0.12 & 1.27 & 1.49 & -0.56 $\pm$ 0.32 & -4.14 $\pm$ 0.39 & U & 85 & 56 \\
        \object{NGC~6793} & 290.824 & 22.140 & 1.66 $\pm$ 0.16 & 0.91 & 1.38 & 3.77 $\pm$ 0.30 & 3.51 $\pm$ 0.26 & T & 60 & 27 \\
        \object{NGC~6811} & 294.342 & 46.395 & 0.98 $\pm$ 0.11 & 0.82 & 0.81 & -4.60 $\pm$ 0.18 & -7.69 $\pm$ 0.17 & U & 54 & 27 \\
        \object{NGC~6866} & 300.987 & 44.160 & 0.96 $\pm$ 0.25 & 0.84 & 0.75 & -1.90 $\pm$ 0.63 & -4.97 $\pm$ 0.27 & U & 9 & 4 \\
        \object{NGC~6913} & 306.000 & 38.510 & 0.52 $\pm$ 0.06 & 0.65 & 0.85 & -3.97 $\pm$ 0.13 & -5.51 $\pm$ 0.13 & U & 58 & 51 \\
        \object{NGC~6940} & 308.610 & 28.280 & 0.93 $\pm$ 0.05 & 1.3 & 1.18 & -2.10 $\pm$ 0.11 & -9.45 $\pm$ 0.12 & T & 87 & 74 \\
        \object{NGC~6991} & 313.614 & 47.466 & 1.85 $\pm$ 0.09 & 1.43 & 1.77 & 5.22 $\pm$ 0.16 & 8.79 $\pm$ 0.08 & U & 97 & 42 \\
        \object{NGC~7209} & 331.225 & 46.552 & 1.14 $\pm$ 0.13 & 0.86 & 0.8 & 2.05 $\pm$ 0.19 & 0.38 $\pm$ 0.14 & T & 57 & 28 \\
        \object{NGC~7243} & 333.795 & 49.875 & 1.30 $\pm$ 0.11 & 1.24 & 1.18 & 0.10 $\pm$ 0.10 & -2.33 $\pm$ 0.09 & U & 40 & 29 \\
        \object{Platais~10} & 205.290 & -59.225 & 3.84 $\pm$ 0.18 & 4.07 & 4.07 & -30.22 $\pm$ 0.30 & -10.59 $\pm$ 0.44 & T & 138 & 10 \\
        \object{Platais~3} & 69.450 & 71.470 & 5.60 $\pm$ 0.05 & 5.0 & 5.88 & 3.51 $\pm$ 0.26 & -21.10 $\pm$ 0.20 & T & 343 & 15 \\
        \object{Platais~8} & 136.875 & -59.160 & 7.59 $\pm$ 0.11 & 7.58 & 7.09 & -15.82 $\pm$ 0.29 & 14.93 $\pm$ 0.32 & T & 457 & 19 \\
        \object{Platais~9} & 137.955 & -43.530 & 5.59 $\pm$ 0.26 & 5.75 & 5.0 & -25.47 $\pm$ 0.38 & 13.54 $\pm$ 0.30 & T & 342 & 21 \\
        \object{Roslund~3} & 299.662 & 20.514 & 0.77 $\pm$ 0.15 & 0.68 & 0.63 & -1.36 $\pm$ 0.32 & -4.18 $\pm$ 0.26 & U & 19 & 14 \\
        \object{Roslund~6} & 307.140 & 39.205 & 2.68 $\pm$ 0.10 & 2.22 & 1.84 & 5.18 $\pm$ 0.21 & 1.75 $\pm$ 0.20 & U & 152 & 41 \\
        \object{Ruprecht~145} & 282.660 & -18.227 & 1.47 $\pm$ 0.08 & 3.13 & 1.3 & 7.51 $\pm$ 0.33 & -2.60 $\pm$ 0.23 & U & 85 & 30 \\
        \object{Ruprecht~147} & 289.092 & -16.250 & 3.26 $\pm$ 0.09 & 3.39 & 3.7 & -1.48 $\pm$ 0.26 & -26.85 $\pm$ 0.20 & U & 194 & 63 \\
        \object{Ruprecht~1} & 99.135 & -14.193 & 0.87 $\pm$ 0.15 & 0.68 & 0.83 & -1.26 $\pm$ 0.30 & 0.16 $\pm$ 0.18 & T & 28 & 16 \\
        \object{Ruprecht~98} & 179.715 & -64.588 & 2.08 $\pm$ 0.10 & 2.02 & 1.64 & -4.30 $\pm$ 0.16 & -8.66 $\pm$ 0.13 & T & 138 & 20 \\
        \object{Stock~10} & 84.810 & 37.805 & 2.90 $\pm$ 0.20 & 2.63 & 2.0 & -4.16 $\pm$ 0.27 & -1.08 $\pm$ 0.29 & U & 87 & 13 \\
        \object{Stock~12} & 353.895 & 52.685 & 2.17 $\pm$ 0.10 & 2.08 & 2.33 & 8.22 $\pm$ 0.23 & -2.32 $\pm$ 0.13 & U & 84 & 22 \\
        \object{Stock~1} & 293.955 & 25.175 & 2.31 $\pm$ 0.05 & 3.14 & 2.86 & 5.92 $\pm$ 0.09 & 0.49 $\pm$ 0.12 & T & 213 & 50 \\
        \object{Stock~2} & 33.660 & 59.440 & 2.78 $\pm$ 0.03 & 3.3 & 2.5 & 16.26 $\pm$ 0.06 & -13.79 $\pm$ 0.05 & U & 226 & 149 \\
        \object{Stock~7} & 37.432 & 60.675 & 1.44 $\pm$ 0.26 & 1.43 & 1.18 & -4.57 $\pm$ 0.19 & 1.11 $\pm$ 0.16 & U & 48 & 12 \\
        \object{Trumpler~10} & 131.872 & -42.400 & 2.28 $\pm$ 0.07 & 2.36 & 2.4 & -12.38 $\pm$ 0.13 & 6.84 $\pm$ 0.12 & T & 161 & 59 \\
        \object{Trumpler~2} & 39.367 & 55.950 & 1.49 $\pm$ 0.10 & 1.38 & 1.49 & 0.14 $\pm$ 0.16 & -6.28 $\pm$ 0.13 & U & 50 & 14 \\
        \object{Trumpler~33} & 276.172 & -19.720 & 0.59 $\pm$ 0.13 & 0.57 & 0.78 & -0.81 $\pm$ 0.76 & -1.01 $\pm$ 1.41 & U & 10 & 7 \\
        \object{Trumpler~3} & 48.052 & 63.150 & 1.36 $\pm$ 0.10 & 2.17 & 1.61 & -3.90 $\pm$ 0.18 & -0.25 $\pm$ 0.10 & U & 46 & 20 \\
        \object{Turner~5} & 143.317 & -36.410 & 2.49 $\pm$ 0.08 & 2.5 & 2.5 & 0.19 $\pm$ 0.29 & -2.67 $\pm$ 0.24 & T & 83 & 11 \\
        \object{vdBergh~92} & 106.020 & -11.530 & 0.86 $\pm$ 0.21 & 0.67 & 0.64 & -5.28 $\pm$ 0.41 & 2.29 $\pm$ 0.30 & T & 20 & 10 \\
        \hline
        \end{tabular}
\tablefoot{Flag $\dagger$U indicates that the UCAC4 proper motions were used, T indicates TGAS proper motions. $\varpi_{\textrm{DAML}}$ and $\varpi_{\textrm{MWSC}}$ are the expected parallaxes assuming no error on the catalogue value. }
\end{center}
\end{table*}

The astrometric validation of GDR1 revealed local systematic biases of the order of 0.3 mas in the parallax zero-point \citep{Arenou17}. Although this does not affect our ability to distinguish between background and foreground stars within a small field of view, the uncertainty on the absolute parallax of a given group of stars should include an additional 0.3\,mas, which cannot be reduced by averaging over a large number of cluster members. 

We compared the mean cluster parallaxes found in this study with the expected values from the DAML and MWSC catalogues. We found two outliers (Collinder~359 and Ruprecht~145) for which the mean parallax is significantly smaller (by more than 3-$\sigma$) than the expected value in DAML, but which are in excellent agreement with the value listed in MWSC. Similarly, one cluster (ASCC~99) has a mean parallax higher by more than 4-$\sigma$ of  the MWSC-expected value, but which is in excellent agreement with DAML. Apart from those three outliers, the standard deviation of the difference between mean and expected prallax in units of uncertainty is 0.75 using DAML references and 0.71 using MWSC. The fact that the values derived from TGAS data differ on average by less than one unit of uncertainty with the catalogue value suggests that, in most cases, adding 0.3\,mas to the total error budget leads to overestimated uncertainties.

The studies of \citet{Stassun16} and \citet{Jao16} report that the TGAS parallaxes could be underestimated by 0.25\,mas, although they note that the effect is larger for stars within 25\,pc and seems to vanish when $\varpi>1$\,mas.
In this present study, the median difference between the mean parallax and the parallax expected from the literature distance is $-0.035$\,mas when using DAML as reference (M.A.D. 0.24) and $-0.032$\,mas (M.A.D. 0.25) using MWSC. These numbers are in agreement with a zero-point of $-0.04\pm0.003$\,mas reported in \citet{Arenou17}, and with the results of \citet{Casertano17} and \citet{Sesar17}, who find no significant zero-point offset of the parallaxes in the TGAS data.

\subsection{Comparison with other results from TGAS parallaxes}

We tested the capability of our approach to reproduce the results obtained in \citet{FvL17} with TGAS parallaxes, for the six most distant OCs in their list ($\varpi<3$\,mas). The mean parallaxes we obtain are in perfect agreement for four of those OCs (see Table~\ref{tab:comparFvL}), but differ for the other two (NGC~2516 and NGC~3532), despite our method providing a list of members very similar to that listed in \citet{FvL17}. The origin of this discrepancy is still under investigation, but we remark that very strong correlations are present for the astrometric parameters of the sources in the field of these two objects.

\begin{table}
\begin{center}
        \caption{Mean parallaxes for the most distant clusters in FvL17. \label{tab:comparFvL}}
        \small\addtolength{\tabcolsep}{-1pt}
        \begin{tabular}{l l l}
        \hline
        \hline
        OC      &       $\langle \varpi \rangle$  &     $\langle \varpi \rangle$ \\
                & \citet{FvL17}                   &     (this study)    \\
        \hline
        \object{NGC~6633}        &       2.41 $\pm$ 0.04 & 2.38 $\pm$ 0.03       \\
        \object{NGC~3532}        &       2.42 $\pm$ 0.04 & 2.26 $\pm$ 0.05       \\
        \object{NGC~2547}        &       2.75 $\pm$ 0.08 & 2.73 $\pm$ 0.10       \\
        \object{IC~4665}         &       2.83 $\pm$ 0.05 & 2.71 $\pm$ 0.08       \\
        \object{Collinder~140}   &       2.86 $\pm$ 0.11 & 2.72 $\pm$ 0.09       \\
        \object{NGC~2516}        &       2.99 $\pm$ 0.08 & 2.51 $\pm$ 0.05       \\
        \hline
        
        \end{tabular}
\tablefoot{Mean parallax $\langle \varpi \rangle$ expressed in mas.}
\end{center}
\end{table}

\subsection{Remarks on individual clusters}

\subsubsection*{Chereul~1}
The putative object listed in \citet{Chereul99} can be clearly identified in proper motion space from TGAS data. Given its estimated distance of $\sim100$\,pc, a better calculation of its mean astrometric parameters taking into account the projection effects and variation of proper motions across the sky \citep[such as in e.g.][]{FvL17} would provide a more accurate determination.
We, however, failed to identify the groups Chereul~2 and Chereul~3, which could be asterisms, and should be further investigated with the next \textit{Gaia} data release.

\subsubsection*{The overlapping objects NGC~1746, NGC~1750, and NGC~1758}
Although they were historically listed as three distinct objects, the reality of all \object{NGC~1746}, \object{NGC~1750}, and \object{NGC~1758} has been questioned by different authors. In particular, \cite{Galadi98i,Galadi98ii} have suggested that NGC~1750 and NGC~1758 might be true physical objects, while NGC~1746 could be an asterism. The more recent study of \citet{Landolt10} was not able to solve the issue of the reality of all three objects. The catalogue of DAML lists them as three different components, with distinct proper motions and distances (800\,pc, 630\,pc, and 760\,pc, respectively) while MWSC lists the whole complex as NGC~1746. In this paper we only identified the system as one group, whose estimated distance and proper motions are closest to those of NGC~1750 in DAML, and therefore list our result under that name.

\section{Age determination from photometry: Bayesian isochrone fitting} \label{sec:base9}

We determined cluster parameters for our clusters with the freely available code Bayesian Analysis for Stellar
Evolution with Nine Parameters \citep[BASE-9][]{vonHippel06}. The code derives posterior distributions for cluster parameters using a Bayesian approach and performing sampling in parameter space with a Markov Chain Monte Carlo technique (MCMC). Given a set of stellar isochrones, BASE-9 compares the theoretical photometry with the observed one, taking into account the presence of unresolved binary stars as well as the probability that each individual star is a cluster member or a field star. More insight on the capabilities of BASE-9 can be found in \citet{Jeffery16}. We set BASE-9 to provide a posterior distribution for four parameters: age ($\log t$), metallicity ([Fe/H]), distance modulus ($(m-M)_V$), and $V$-band extinction ($A_V=3.1 \times E(B-V)$). We made use of PARSEC isochrones \citep{Bressan12}, which are not shipped with BASE-9 but were straightforward to implement.

Although the $G$-magnitudes contained in the GDR1 catalogue are of exquisite quality, we decided not to make use of them because the common approximation that all stars are affected equally by interstellar extinction does not hold for such precise photometric measurements. 
In the presence of interstellar extinction, it is common to assume that all stars are affected in the same way regardless of their spectral type, and that absorption produces an identical, rigid shift in colour and magnitude for all stars in a cluster. In reality, stars of different spectral types are affected differently by extinction \citep{Jordi10}, with variations of up to 10--15\,mmag in $G$-band, even in cases of moderate extinction with $A_V\sim0.5$ (Sordo et al., in prep.). This effect must be accounted for when working with \textit{Gaia} photometry, which for the subset of TGAS stars (brighter than $G=13$) has a median photometric error of 1.3\,mmag. Adapting BASE-9 to the specific requirements of sub-millimag photometry is beyond the scope of this study, and we limited the analysis to 2MASS $JHK_S$ photometry using the pre-computed cross-match between the \textit{Gaia} and 2MASS catalogues provided with the GDR1 (Marrese et al. in prep.).

A Bayesian approach requires the setting of priors on the parameters we are trying to determine and BASE-9 only allows for either flat or Gaussian priors. We set a Gaussian prior on distance modulus, centred on the value found in Sect.~\ref{sec:meanparams} using the mean parallax.
We set rather loose Gaussian priors on the other three parameters, with $\log t$ centred on the value listed in the MWSC catalogue and with a dispersion of 0.5. The prior on the extinction $A_V$ was also centred on the literature value, with a dispersion of 0.2. 
The metallicity prior distribution was centred on [Fe/H]=-0.1 for all stars, with a dispersion of 0.2. When spectroscopic metallicities were available, we centred the prior on that value, with a dispersion of 0.2\,dex as well. BASE-9 also allows us to take into account prior knowledge of membership probability, for which we used the results obtained in Sect.~\ref{sec:membership}.

Following the approach of \citet{Jeffery16}, for every OC we performed five runs of BASE-9 sampling 3000 points each, using slightly different starting points. The first run started from the literature value for age, extinction, and metallicity, and the distance modulus used was the one determined in Sect.~\ref{sec:meanparams}. The starting points for the additional four runs were identical, but we shifted either the initial distance modulus by $\pm0.3$ or the age by $\pm0.2$ in logarithmic scale.
These values correspond to the average accuracy in the MWSC Catalogue. 

In most cases the output of BASE-9 proved sensitive to the starting values, which we attribute to the low number of member stars in most of our clusters. When the inclusion or rejection of one star can have a strong influence on the choice of the best fit isochrone, the likelihood space can have a chaotic structure with deep local minima,
and the choice of initial point for the sampling can have a strong influence on the final results.
For 26 OCs, the procedure converged to similar posterior distributions in all five runs and provided a satisfactory fit to the observed colour-magnitude diagram (CMD). 
Unsurprisingly, these clusters tend to be those with a larger number of members or featuring red clump stars, which provide good constrains on the cluster parameters.

For those 26 OCs, we combined the results of all five runs to compute the final cluster parameters and their uncertainty (effectively using a total of 15\,000 samplings). 
The posterior distributions are generally non-symmetrical and show correlations between parameters (e.g. age and extinction, see Fig.~\ref{fig:NGC_2567_posteriormap}). For simplicity we only report here the mean and standard deviation of those posterior distributions. As pointed out by \citet{Jeffery16}, the uncertainties on the cluster parameters reflect the internal precision of the procedure (the certainty with which a certain PARSEC isochrone represents the data better than a different PARSEC isochrone), rather than an absolute accuracy. An example of CMD, along with isochrones sampled from the posterior distribution, is shown in Fig.~\ref{fig:isochrones_bestfit_ngc2567}.

\begin{figure}[ht]
\begin{center} \resizebox{\hsize}{!}{\includegraphics[scale=0.5]{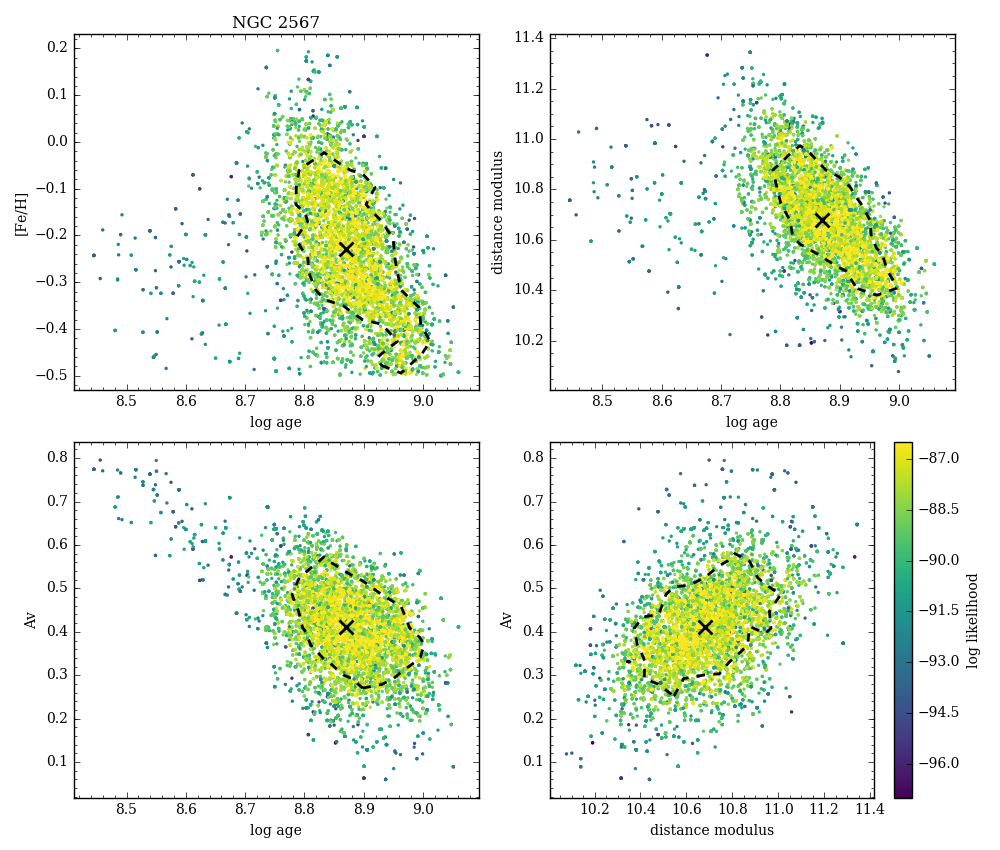}} \caption{\label{fig:NGC_2567_posteriormap} Posterior distribution maps for NGC~2567 combining the outputs of five BASE-9 runs. The dashed contour encircles 68\% of the total likelihood and the crossed symbol shows the mean value. } \end{center}
\end{figure}

\begin{figure}[ht]
\begin{center} \resizebox{\hsize}{!}{\includegraphics[scale=0.5]{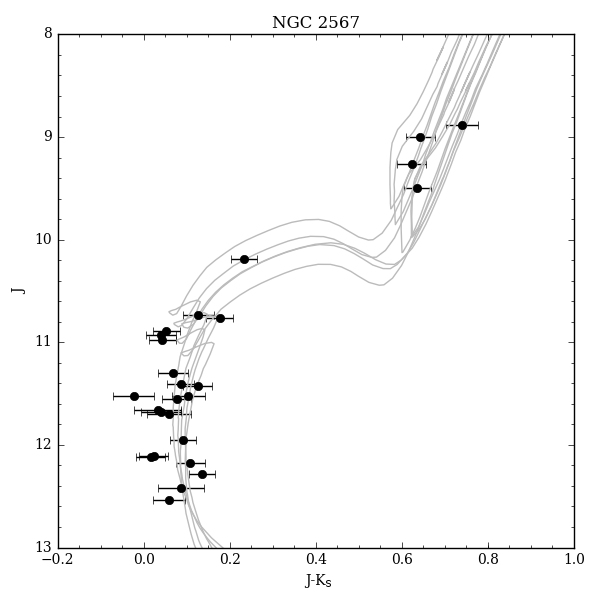}} \caption{\label{fig:isochrones_bestfit_ngc2567} Colour-magnitude diagram for NGC~2567. The grey lines are five PARSEC ischrones randomly chosen from the posterior distribution returned by BASE-9.} \end{center}
\end{figure}

\begin{figure}[ht]
\begin{center} \resizebox{\hsize}{!}{\includegraphics[scale=0.5]{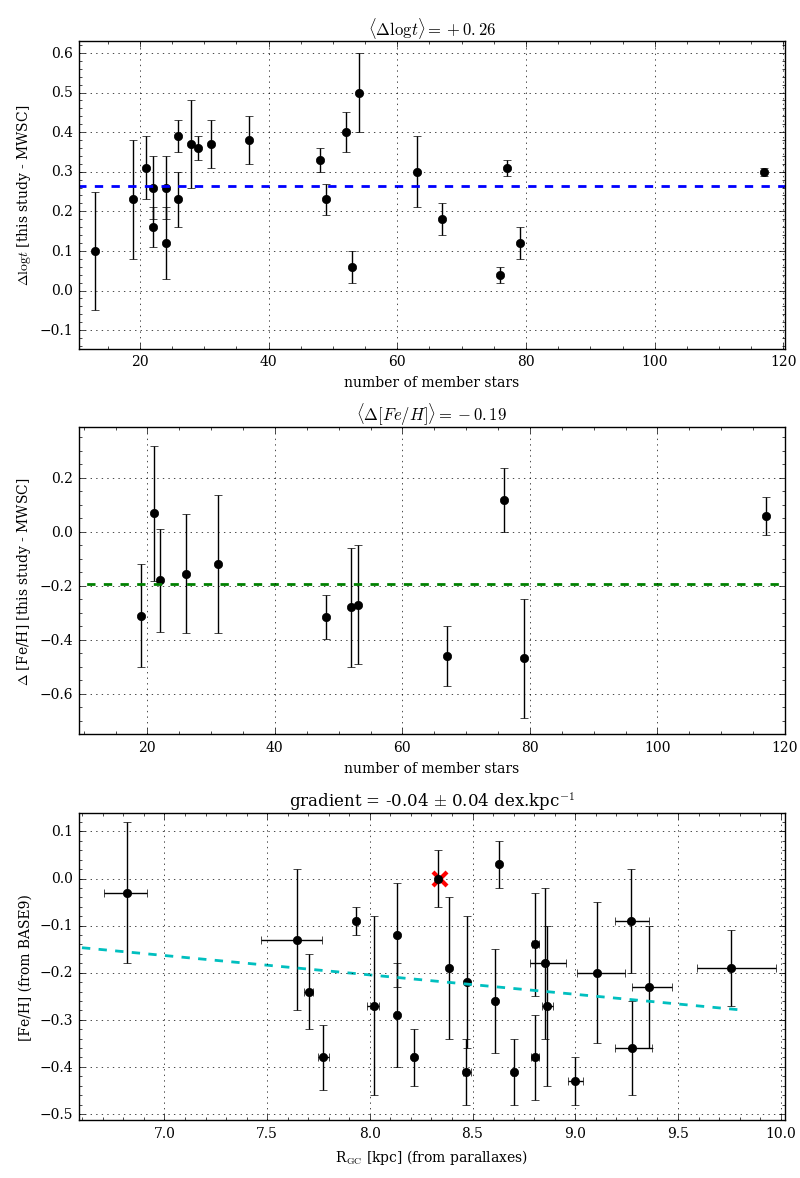}} \caption{\label{fig:stats_base9} Top: Difference in $\log t$ between the ages found in this study and those quoted in the MWSC catalogue. The dashed line indicates the mean value. Middle: Difference between the [Fe/H] found in this study and those quoted in the MWSC catalogue (for the OCs with metallicity estimates). The dashed blue line indicates the mean value. Bottom: Metallicity ([Fe/H]) obtained in this study based on $JHK_S$ photometry, against Galactocentric radius. The red cross indicates the solar metallicity and Galactocentric radius \citep[8.34\,kpc,][]{Reid14}.} \end{center}
\end{figure}

Our age ($\log t$) determinations are on average older by 0.26 than the literature values listed in the MWSC.
Systematic differences are expected between studies that make use of different sets of models, and the MWSC ages \citep{Kharchenko13} were determined from PADOVA isochrones, while this study makes use of PARSEC models. In particular, different choices of solar metallicity reference and mixing-length parameter are known to lead to slight differences in the predicted brightness of red clump stars, which in turn has an incidence on the choice of model that best reproduces the morphology of the CMD \citep[see e.g. Section 7.1 in][]{CantatGaudin14m11}.

Metallicity is the parameter that is least well constrained by isochrone fitting. The mean metallicity derived from photometry for this sample of OCs is [Fe/H]=-0.23\,dex. We find that our determination of the metallicity is systematically lower by $\Delta$[Fe/H]$=0.19$ in comparison to  MWSC values. Although a significant scatter is present, our sample traces a marginally negative metallicity gradient, with slope -0.04$\pm$0.04\,dex\,kpc$^{-1}$, as shown in the bottom panel of Fig.~\ref{fig:stats_base9}. The uncertainty on the gradient slope was computed as the standard deviation among 1000 redrawings, performed by picking a metallicity and a mean parallax from a Gaussian distribution representing the values found in this study and their associated errors.
This value is compatible with the results of \citet{Netopil16} or \citet{Jacobson16} (-0.10$\pm$0.02 and -0.085$\pm$0.017\,dex\,kpc$^{-1}$, respectively). Our sample is, however, rather small (26 objects) and the precision of  metallicites derived from isochrones fitting is not as good as the results that can be obtained from high-resolution spectroscopy.

\begin{table}
\begin{center}
        \caption{ \label{tab:meanbase9params} Cluster parameters derived from $JHK_S$ photometry for 26 OCs.}
        \small\addtolength{\tabcolsep}{-1pt}
        \begin{tabular}{l l l l l}
        \hline
        \hline
   OC   & $\log t$ & [Fe/H] & $A_V$ & dist. mod.      \\
        \hline

Alessi~2 &    8.96 $\pm{0.08}$    & -0.27 $\pm{0.17}$    & 0.45 $\pm{0.13}$   & 8.52 $\pm{0.20}$ \\
IC~4756 &    9.09 $\pm{0.01}$    & -0.09 $\pm{0.03}$    & 0.31 $\pm{0.04}$   & 8.24 $\pm{0.05}$ \\
NGC~0752 &    9.17 $\pm{0.02}$    & 0.03 $\pm{0.05}$    & 0.11 $\pm{0.05}$   & 8.19 $\pm{0.07}$ \\
NGC~1528 &    8.94 $\pm{0.04}$    & -0.09 $\pm{0.11}$    & 0.30 $\pm{0.10}$   & 9.49 $\pm{0.14}$ \\
NGC~1662 &    9.03 $\pm{0.03}$    & -0.41 $\pm{0.07}$    & 0.66 $\pm{0.06}$   & 7.60 $\pm{0.08}$ \\
NGC~1750 &    8.58 $\pm{0.09}$    & -0.43 $\pm{0.05}$    & 0.94 $\pm{0.07}$   & 9.09 $\pm{0.09}$ \\
NGC~1912 &    8.85 $\pm{0.10}$    & -0.36 $\pm{0.10}$    & 0.69 $\pm{0.10}$   & 9.77 $\pm{0.13}$ \\
NGC~2099 &    8.95 $\pm{0.05}$    & -0.19 $\pm{0.08}$    & 0.50 $\pm{0.08}$   & 9.69 $\pm{0.14}$ \\
NGC~2281 &    8.85 $\pm{0.04}$    & -0.14 $\pm{0.11}$    & 0.29 $\pm{0.06}$   & 8.40 $\pm{0.12}$ \\
NGC~2482 &    8.88 $\pm{0.06}$    & -0.20 $\pm{0.15}$    & 0.15 $\pm{0.08}$   & 9.76 $\pm{0.16}$ \\
NGC~2527 &    9.03 $\pm{0.04}$    & -0.26 $\pm{0.11}$    & 0.24 $\pm{0.05}$   & 8.73 $\pm{0.10}$ \\
NGC~2539 &    8.96 $\pm{0.15}$    & -0.18 $\pm{0.16}$    & 0.29 $\pm{0.15}$   & 9.97 $\pm{0.22}$ \\
NGC~2548 &    8.90 $\pm{0.04}$    & -0.38 $\pm{0.09}$    & 0.28 $\pm{0.05}$   & 9.07 $\pm{0.08}$ \\
NGC~2567 &    8.87 $\pm{0.07}$    & -0.23 $\pm{0.13}$    & 0.41 $\pm{0.09}$   & 10.68 $\pm{0.18}$ \\
NGC~4852 &    8.89 $\pm{0.08}$    & -0.13 $\pm{0.15}$    & 0.26 $\pm{0.12}$   & 9.37 $\pm{0.19}$ \\
NGC~5822 &    9.15 $\pm{0.02}$    & -0.24 $\pm{0.08}$    & 0.41 $\pm{0.10}$   & 9.32 $\pm{0.09}$ \\
NGC~6152 &    8.75 $\pm{0.09}$    & -0.03 $\pm{0.15}$    & 0.69 $\pm{0.14}$   & 10.34 $\pm{0.27}$ \\
NGC~6281 &    8.80 $\pm{0.09}$    & -0.38 $\pm{0.07}$    & 0.63 $\pm{0.07}$   & 8.19 $\pm{0.11}$ \\
NGC~6793 &    9.07 $\pm{0.11}$    & -0.27 $\pm{0.19}$    & 0.47 $\pm{0.10}$   & 8.56 $\pm{0.20}$ \\
NGC~6811 &    9.16 $\pm{0.03}$    & -0.38 $\pm{0.06}$    & 0.36 $\pm{0.07}$   & 9.75 $\pm{0.10}$ \\
NGC~6991 &    9.15 $\pm{0.02}$    & -0.00 $\pm{0.06}$    & 0.25 $\pm{0.06}$   & 8.82 $\pm{0.11}$ \\
NGC~7209 &    9.01 $\pm{0.06}$    & -0.41 $\pm{0.07}$    & 0.34 $\pm{0.10}$   & 9.45 $\pm{0.15}$ \\
Platais~3 &    8.90 $\pm{0.15}$    & -0.22 $\pm{0.14}$    & 0.17 $\pm{0.08}$   & 6.23 $\pm{0.11}$ \\
Ruprecht~98 &    8.96 $\pm{0.05}$    & -0.29 $\pm{0.11}$    & 0.53 $\pm{0.09}$   & 8.01 $\pm{0.12}$ \\
Stock~1 &    8.77 $\pm{0.04}$    & -0.12 $\pm{0.11}$    & 0.27 $\pm{0.07}$   & 7.76 $\pm{0.10}$ \\
Turner~5 &    8.80 $\pm{0.08}$    & -0.19 $\pm{0.15}$    & 0.11 $\pm{0.08}$   & 7.58 $\pm{0.17}$ \\

        \hline
        \end{tabular}
\end{center}
\end{table}

\section{Computing three-dimensional velocities and full orbits} \label{sec:orbits}
In order to compute full, three-dimensional velocities, we combined the proper motions determined in this study with spectroscopic radial velocities listed in \citet{Mermilliod08} and \citet{Mermilliod09}. After excluding the non-members and the stars flagged as either variables or binaries, we found that 36 of the OCs in our sample have radial velocities that can be computed from at least two stars.

The velocity of a particle in the Galactic disk can be described as the sum of two components: i) a circular motion around the Galactic centre, at a velocity depending on the Galactocentric radius and which defines the Regional Standard of Rest (RSR), and ii) an additional peculiar motion with respect to this RSR. We computed the cylindrical $(U_S,V_S,W_S)$\footnote{The velocity component $U_S$ is positive towards the Galactic centre, $V_S$ along the Galactic rotation, and $W_S$ towards the Galactic north pole. } components of the peculiar motion as described in Sect.~3.5.1 of \citet{Casamiquela16}, adopting a Galactic rotation curve with $\Theta_0=240$\,km\,s$^{-1}$, $R_{\odot}=8.34$\,kpc, and $\frac{dR}{d\Theta}=-0.2$\,km\,s$^{-1}$\,kpc$^{-1}$, and taking into account the Sun's own peculiar motion as $(U_{\odot},V_{\odot},W_{\odot})=(10.7,15.6,8.9)$\,km\,s$^{-1}$ \citep{Reid14}. The results of this computation are listed in Table~\ref{tab:orbits} and shown in Fig.~\ref{fig:UVW}. We found velocity dispersions of 21.1, 14.8, and 8.8\,km\,s$^{-1}$ for $U_S$, $V_S$, and $W_S$ (respectively). These values are in good agreement with the values of 20, 15, and 10\,km\,s$^{-1}$ found by \citet{Holmberg09} for stars younger than 2\,Gyr.

\begin{figure*}[ht]
\begin{center} \resizebox{\hsize}{!}{\includegraphics[scale=0.5]{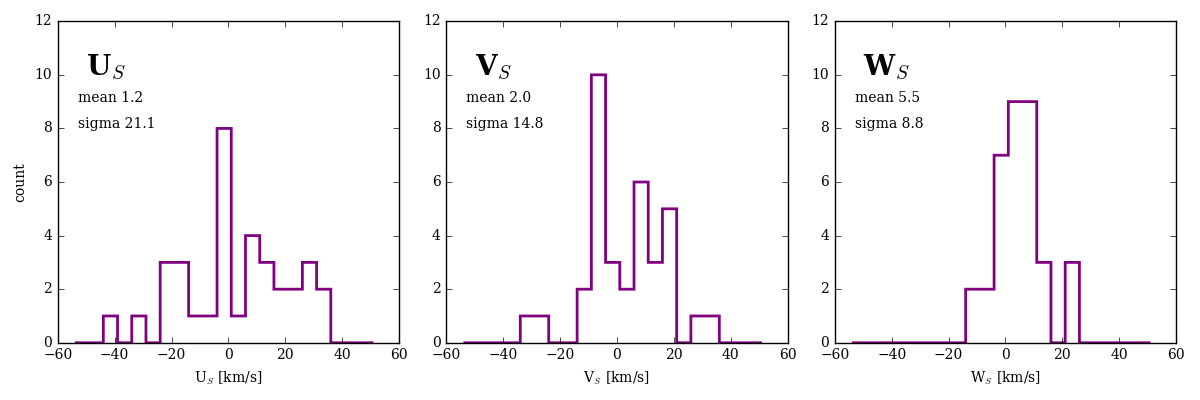}} \caption{\label{fig:UVW} Distribution of the three components of the peculiar motion for the OCs listed in Table~\ref{tab:orbits}. NGC~5617 was discarded because of the large uncertainty affecting its distance determination.  } \end{center}
\end{figure*}

From the three-dimensional position and three-dimensional velocity of each cluster, we computed orbits using the software \texttt{galpy} and the static, axisymmetric potential \texttt{MWPotential2014} \citep{Bovy15}. Figure~\ref{fig:zmax} shows the current distance ($|z|$) and maximum altitude above the Galactic plane ($z_{\mathrm{max}}$) for those integrated orbits. Both these quantities correlate with age, with OCs younger than 300\,Myr being all contained within 180\,pc of the plane, while half the older clusters have orbits that extend beyond this limit. The oldest OC in our sample is NGC~2682, one of the oldest known open clusters, and its orbit strays more than 400\,pc from the Galactic plane. We also computed the eccentricity of each orbit ($e=\frac{r_a-r_p}{r_a+r_p}$ where $r_p$ and $r_a$ are the perigalacticon and apogalacticon of the orbit), and found no apparent correlation of eccentricity with age (bottom panel of Fig.~\ref{fig:zmax}).
The orbital parameters of the integrated orbits are listed in Table~\ref{tab:orbits}.

\begin{figure}[ht]
\begin{center} \resizebox{\hsize}{!}{\includegraphics[scale=0.5]{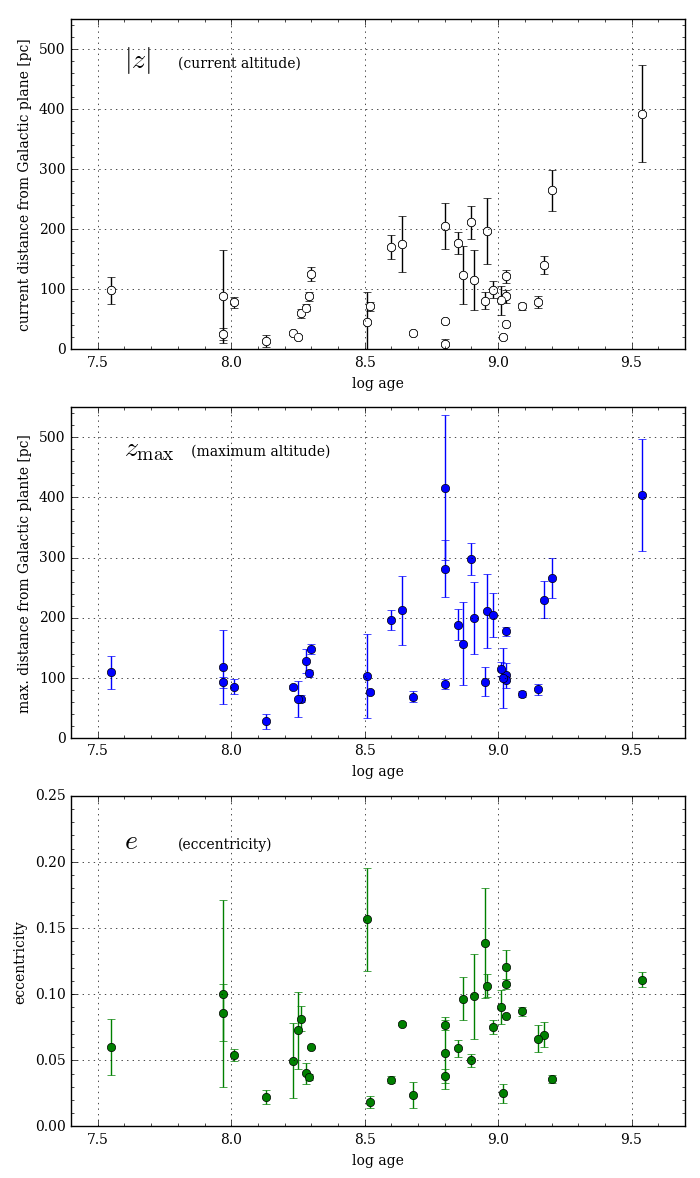}} \caption{\label{fig:zmax} Top: Current distance $|z|$ from the Galactic plane as a function of age for the 36 OCs for which we computed full orbits. Middle: Maximum altitude above the Galactic plane for the integrated orbits of those OCs. Bottom: Eccentricity of the integrated orbits against age of the cluster.} \end{center}
\end{figure}

\begin{table*}
\begin{center}
        \caption{ \label{tab:orbits} Selected parameters for 36 integrated orbits.}
        \small\addtolength{\tabcolsep}{-2pt}
        \begin{tabular}{l l l l l l l l}
        \hline
        \hline
   OC   & Log age & $U_S$ & $V_S$ & $W_S$ & $z$  & $z_{\mathrm{max}}$ & e    \\
        &         &       &       &       & [pc] & [pc]               &      \\
        \hline

IC~4725 & 7.97 & 13.7 $\pm$ 0.5 & -4.4 $\pm$ 1.4 & 10.4 $\pm$ 1.1 & -25 $\pm$ 9 & 92 $\pm$ 9 & 0.086 $\pm$ 0.021 \\
IC~4756 & 9.09 & -15.3 $\pm$ 0.3 & -3.4 $\pm$ 0.3 & 2.3 $\pm$ 0.3 & 72 $\pm$ 5 & 72 $\pm$ 5 & 0.087 $\pm$ 0.004 \\
NGC~0752 & 9.17 & -13.1 $\pm$ 0.3 & -6.0 $\pm$ 0.5 & -9.5 $\pm$ 0.3 & -141 $\pm$ 14 & 229 $\pm$ 30 & 0.069 $\pm$ 0.009 \\
NGC~1342 & 8.6 & 11.9 $\pm$ 0.4 & 8.5 $\pm$ 0.6 & 8.1 $\pm$ 0.5 & -170 $\pm$ 19 & 196 $\pm$ 17 & 0.035 $\pm$ 0.003 \\
NGC~1647 & 8.3 & 18.8 $\pm$ 0.3 & 14.0 $\pm$ 0.3 & 6.5 $\pm$ 0.3 & -125 $\pm$ 11 & 148 $\pm$ 8 & 0.06 $\pm$ 0.0 \\
NGC~1662 & 9.03 & 26.1 $\pm$ 0.4 & 15.5 $\pm$ 0.2 & 10.2 $\pm$ 0.3 & -121 $\pm$ 10 & 177 $\pm$ 7 & 0.084 $\pm$ 0.001 \\
NGC~2099 & 8.95 & -1.1 $\pm$ 0.3 & -30.6 $\pm$ 4.0 & 1.2 $\pm$ 1.8 & 81 $\pm$ 13 & 93 $\pm$ 24 & 0.138 $\pm$ 0.042 \\
NGC~2168 & 8.26 & 23.1 $\pm$ 0.3 & 0.1 $\pm$ 0.6 & 2.8 $\pm$ 0.5 & 59 $\pm$ 7 & 64 $\pm$ 5 & 0.081 $\pm$ 0.01 \\
NGC~2281 & 8.85 & -15.1 $\pm$ 0.2 & 4.8 $\pm$ 0.5 & -3.1 $\pm$ 0.6 & 176 $\pm$ 18 & 188 $\pm$ 25 & 0.059 $\pm$ 0.006 \\
NGC~2360 & 8.8 & -4.8 $\pm$ 1.1 & 17.9 $\pm$ 1.1 & 24.6 $\pm$ 1.2 & -9 $\pm$ 8 & 415 $\pm$ 120 & 0.055 $\pm$ 0.027 \\
NGC~2423 & 9.03 & 26.4 $\pm$ 0.5 & -7.9 $\pm$ 0.5 & -0.8 $\pm$ 0.6 & 88 $\pm$ 10 & 104 $\pm$ 20 & 0.12 $\pm$ 0.013 \\
NGC~2447 & 8.68 & -1.4 $\pm$ 0.8 & 10.8 $\pm$ 0.6 & 5.6 $\pm$ 0.6 & 27 $\pm$ 0 & 68 $\pm$ 9 & 0.023 $\pm$ 0.01 \\
NGC~2477 & 8.91 & 35.7 $\pm$ 4.4 & 10.7 $\pm$ 2.4 & 9.6 $\pm$ 5.2 & -115 $\pm$ 49 & 198 $\pm$ 59 & 0.098 $\pm$ 0.032 \\
NGC~2527 & 9.03 & -15.1 $\pm$ 0.3 & -8.6 $\pm$ 0.2 & 6.8 $\pm$ 0.2 & 42 $\pm$ 1 & 97 $\pm$ 1 & 0.107 $\pm$ 0.004 \\
NGC~2539 & 8.96 & 8.3 $\pm$ 1.0 & -9.3 $\pm$ 0.9 & 2.5 $\pm$ 1.7 & 196 $\pm$ 54 & 211 $\pm$ 62 & 0.106 $\pm$ 0.009 \\
NGC~2546 & 8.13 & 4.2 $\pm$ 1.3 & 10.9 $\pm$ 0.8 & -1.9 $\pm$ 0.8 & -13 $\pm$ 9 & 27 $\pm$ 12 & 0.022 $\pm$ 0.005 \\
NGC~2548 & 8.9 & 14.4 $\pm$ 0.6 & 15.8 $\pm$ 0.6 & 14.2 $\pm$ 0.7 & 211 $\pm$ 27 & 297 $\pm$ 27 & 0.05 $\pm$ 0.005 \\
NGC~2567 & 8.87 & 10.4 $\pm$ 1.7 & -5.5 $\pm$ 1.0 & -2.3 $\pm$ 1.2 & 123 $\pm$ 48 & 156 $\pm$ 68 & 0.097 $\pm$ 0.016 \\
NGC~2682 & 9.54 & -19.0 $\pm$ 1.7 & -11.0 $\pm$ 1.3 & -3.0 $\pm$ 2.9 & 392 $\pm$ 80 & 403 $\pm$ 93 & 0.111 $\pm$ 0.006 \\
NGC~3680 & 9.2 & -1.8 $\pm$ 2.5 & 1.8 $\pm$ 0.7 & -0.3 $\pm$ 1.0 & 264 $\pm$ 34 & 266 $\pm$ 33 & 0.036 $\pm$ 0.003 \\
NGC~5138 & 7.55 & 0.4 $\pm$ 6.5 & -7.8 $\pm$ 3.5 & 5.2 $\pm$ 1.6 & 98 $\pm$ 21 & 108 $\pm$ 27 & 0.06 $\pm$ 0.021 \\
NGC~5316 & 8.23 & -39.5 $\pm$ 5.0 & -26.0 $\pm$ 3.1 & 21.2 $\pm$ 0.9 & 27 $\pm$ 0 & 84 $\pm$ 3 & 0.05 $\pm$ 0.028 \\
NGC~5617 & 8.25 & -62.5 $\pm$ 20.4 & -35.2 $\pm$ 11.1 & 22.8 $\pm$ 2.9 & 21 $\pm$ 1 & 64 $\pm$ 30 & 0.073 $\pm$ 0.029 \\
NGC~5662 & 8.28 & -22.0 $\pm$ 1.0 & -3.1 $\pm$ 0.9 & 0.9 $\pm$ 0.4 & 69 $\pm$ 5 & 128 $\pm$ 19 & 0.04 $\pm$ 0.008 \\
NGC~5822 & 9.15 & -31.0 $\pm$ 0.8 & -6.2 $\pm$ 0.8 & 13.5 $\pm$ 0.4 & 78 $\pm$ 10 & 81 $\pm$ 9 & 0.066 $\pm$ 0.01 \\
NGC~6067 & 7.97 & -0.7 $\pm$ 4.5 & -7.6 $\pm$ 4.9 & 14.4 $\pm$ 2.5 & -88 $\pm$ 77 & 117 $\pm$ 61 & 0.1 $\pm$ 0.071 \\
NGC~6124 & 8.29 & -3.5 $\pm$ 0.4 & 20.3 $\pm$ 0.7 & 7.0 $\pm$ 0.7 & 88 $\pm$ 7 & 107 $\pm$ 6 & 0.037 $\pm$ 0.002 \\
NGC~6134 & 9.02 & -2.8 $\pm$ 1.1 & 17.8 $\pm$ 2.1 & -8.3 $\pm$ 2.6 & 21 $\pm$ 0 & 99 $\pm$ 50 & 0.025 $\pm$ 0.007 \\
NGC~6281 & 8.8 & 6.9 $\pm$ 0.4 & 6.2 $\pm$ 0.9 & 9.7 $\pm$ 0.9 & 47 $\pm$ 2 & 89 $\pm$ 8 & 0.038 $\pm$ 0.005 \\
NGC~6494 & 8.52 & -2.4 $\pm$ 0.2 & 16.7 $\pm$ 0.9 & 4.5 $\pm$ 1.0 & 71 $\pm$ 7 & 76 $\pm$ 3 & 0.019 $\pm$ 0.005 \\
NGC~6705 & 8.51 & 35.5 $\pm$ 4.7 & -4.2 $\pm$ 7.0 & 4.4 $\pm$ 4.0 & -45 $\pm$ 50 & 102 $\pm$ 70 & 0.157 $\pm$ 0.039 \\
NGC~6811 & 8.8 & 28.3 $\pm$ 3.5 & 16.4 $\pm$ 0.5 & 22.1 $\pm$ 1.4 & 206 $\pm$ 38 & 281 $\pm$ 47 & 0.077 $\pm$ 0.004 \\
NGC~6866 & 8.64 & 7.7 $\pm$ 4.6 & 27.1 $\pm$ 0.6 & 8.8 $\pm$ 3.8 & 175 $\pm$ 46 & 212 $\pm$ 57 & 0.078 $\pm$ 0.001 \\
NGC~6940 & 8.98 & 24.8 $\pm$ 1.6 & 9.5 $\pm$ 0.5 & -10.5 $\pm$ 0.9 & -99 $\pm$ 13 & 205 $\pm$ 36 & 0.075 $\pm$ 0.005 \\
NGC~7209 & 9.01 & -22.4 $\pm$ 1.2 & -5.4 $\pm$ 0.4 & 5.1 $\pm$ 0.9 & -81 $\pm$ 23 & 114 $\pm$ 11 & 0.09 $\pm$ 0.013 \\
Trumpler~3 & 8.01 & 18.2 $\pm$ 1.6 & 31.2 $\pm$ 1.8 & -8.0 $\pm$ 1.1 & 78 $\pm$ 9 & 84 $\pm$ 12 & 0.054 $\pm$ 0.005 \\

        \hline
        \end{tabular}
\tablefoot{The velocity $U_S$ is positive towards the Galactic centre, $V_S$ along the Galactic rotation, and $W_S$ towards the Galactic north pole. The uncertainties were derived from 10\,000 random realisations, varying the cluster distance, proper motion, and radial velocity according to their uncertainties. }
\end{center}
\end{table*}

\section{Discussion} \label{sec:discussion}
The two-million-star TGAS sample allows us to identify members and derive mean parallaxes in clusters located further than 1000\,pc from us, thus enabling studies of the Galactic disk on a larger scale than the works based on the results of the Hipparcos catalogue \citep[e.g.][]{Robichon99,Bouy15}, which is severely incomplete beyond 100\,pc and for which the relative parallax error reaches 50\% at 350\,pc.

The distribution of OCs identified in this paper is shown in Fig.~\ref{fig:map}, along with the schematic location of the spiral arms of the Milky Way in the model of \citet{Reid14}. We remark that almost all OCs in this study are located in the inter-arm region. In order to trace the spiral structure of the thin disk, our sample would require a significant number of objects younger than 20\,Myr ($\log t<7.3$), as clusters drift away from their birthplace over this timescale \citep[see e.g.][]{Dias05}. These young clusters and associations are often sparsely populated and/or embedded in their progenitor molecular cloud (thus requiring the use of infrared photometry). Due to the magnitude limits of the TGAS solution, even nearby clusters such as Chamaeleon~I \citep[2\,Myr old, 160\,pc from us,][]{Whittet97} are not present in the catalogue. Hints of objects such as IC~2395 \citep[6\,Myr, 800\,pc][]{Claria03} can be seen, but not with a sufficient number of stars for our method to establish secure membership.
Another common tracer of the spiral structure of the disk are the so-called OB associations, which are even sparser groups of bright, short-lived blue stars, whose identification requires the use of multi-band photometry (and ideally spectroscopy). Such a search is well beyond the scope of this paper, aimed at identifying stellar clusters from astrometric data.

Out of the 128 OCs for which we derived mean astrometric parameters, 67 have mean proper motions obtained from TGAS data (with a median proper motion error of 0.23\,mas\,yr$^{-1}$), while UCAC4 data was used for 61 of them (these OCs have a median proper motion error of 0.32\,mas\,yr$^{-1}$).

\begin{figure}[ht]
\begin{center} \resizebox{\hsize}{!}{\includegraphics[scale=0.5]{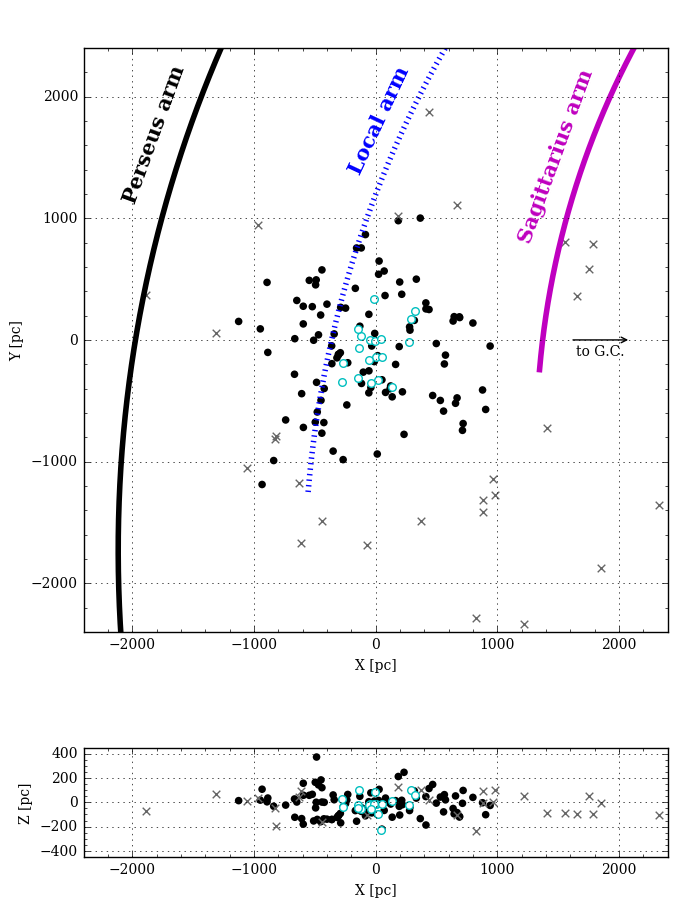}} \caption{\label{fig:map} Position of the OCs studied in this paper in Galactic rectangular $XYZ$ coordinates. Black dots: OCs from this study with relative parallax errors under 50\% (considering an additional 0.3\,mas calibration error for all OCs). Crosses: OCs from this study with relative parallax error over 50\%. Open symbols: OCs from \citet{FvL17}. The spiral arms are traced according to the model of \citet{Reid14}.} \end{center}
\end{figure}

The difference in scale height observed between young and old objects is traditionally attributed to disk heating: stars and stellar clusters are formed on flat and circular orbits, which might be disturbed during the course of their life, in particular by interactions with giant molecular clouds and non-axisymmetric features of the disk \citep{Carlberg87,Aumer16,Grand16}. The clear correlation between age and altitude above the plane shown in Fig.~\ref{fig:zmax}, where clusters younger than 300\,Myr have a mean value of $z_{\mathrm{max}}$ of 100\,pc while the older clusters have a mean $z_{\mathrm{max}}$ of 190\,pc, illustrates the fact that their vertical motion can be affected during the first hundred million years of their life. 

The absence of an apparent correlation between age and eccentricity seems to indicate that radial heating has not affected the objects in our sample as much as vertical heating, and the timescale for radial heating is longer than that of vertical heating. The sample of old outer disk clusters studied by \citet{CantatGaudin16} contains several OCs older than 2\,Gyr and eccentricities larger than 0.2. The study of \citet{VandePutte10} also shows that OCs older than 1\,Gyr tend to have larger eccentricities, but contains a significant number of young objects that appear to follow eccentric orbits as well. Both these studies might be affected by an observational bias, since old stellar clusters are more easily detected at high Galactic latitudes, towards the less crowded anticentre direction, and therefore more likely to be those with perturbed orbits. The uncertainties on proper motions for these distant objects also translate into large uncertainties on their physical motion. 
The unprecedented quality of the \textit{Gaia}-DR2 data will allow for a more accurate characterisation of the kinematics for a much larger sample, giving us a deeper insight into the age-kinematics relations over a large portion of the Galactic disk.

\section{Conclusion and future prospects} \label{sec:conclusion}

In this paper, we make use of a suitable combination of \textit{Gaia} DR1 TGAS parallaxes and proper motions and UCAC4 proper motions to derive a list of high probability cluster members for 128 OCs. For those objects, we compute mean proper motions and parallaxes. For 26 clusters, we obtain parameters such as age, extinction, [Fe/H], and distance modulus from comparison with isochrones  using a Bayesian/MCMC method.

The issue of dealing with a large astronomical dataset is not unique to \textit{Gaia}, as other current \citep[e.g. the Panoramic Survey Telescope and Rapid Response System]{PanSTARRS1} or future \citep[e.g. the Large Synoptic Survey Telescope]{Ivezic08lsst} observational campaigns will deliver multi-dimensional catalogues of unprecedented size.
The work conducted in this study shows the strength of automatic approaches to cluster membership selection such as UPMASK when dealing with large datasets. While the two-million-star TGAS subset of GDR1 is larger than the Hipparcos catalogue by an order of magnitude, and therefore already too rich to be investigated by hand, the second \textit{Gaia} data release (GDR2) will contain a full astrometric solution for one billion star, dwarfing TGAS by a factor 500. 
The upcoming ground-based LSST is expected to reach four magnitudes deeper than \textit{Gaia}, and to deliver better astrometry than \textit{Gaia} for sources fainter than $G\sim20$ as well as multi-band photometry for an end-of-mission catalogue totalling many billions of objects. 

Systematic cluster studies conducted with automated methods should also attempt to provide general cluster parameters such as ages, using stellar evolution models. Ideally, these determinations should be performed for as many clusters as possible in order to build large homogeneous samples and avoid the additional dispersion in results introduced by compiling age determinations originating from various studies making use of various sets of models. Automated tools provide an objective estimate of the cluster parameters, a convenient alternative to fitting colour-magnitude diagrams by eye (which yields non-reproducible results and is completely impractical when dealing with samples of hundreds of clusters), and allows us to consider independent measurements such as distance estimates obtained from trigonometric parallaxes.

This preparatory study shows the strength of our tools, applied to astrometric data, to identify and study stellar clusters.
The upcoming GDR2 dataset will be much deeper than TGAS (with magnitudes reaching down to $G\sim21$, against $G\sim13$ in TGAS), with a better astrometric precision for both proper motions and parallaxes, allowing us to identify and characterise clusters at much larger distances. The data of GDR2 will also contain \textit{Gaia} $G_{BP}$ and $G_{RP}$ magnitudes, allowing us to perform photometric studies using \textit{Gaia} data alone. Efforts should be made towards developing tools to take into account the effect of interstellar extinction on stars of different spectral types (see Sordo et al., in prep.), in order to take full advantage of the sub-millimag precision of \textit{Gaia} magnitudes.

In addition to providing better characterisations of known OCs, the GDR2 data is expected to contain many as yet unknown clusters and associations. Completing the cluster census in the Galactic disk and estimating ages is crucial to understanding cluster formation and disruption, and for tracing the structures and the processes that shape the disk of the Milky Way.

\section*{Acknowledgements}

Part of the work presented in this paper was support by the European Science Foundation under ESF grant number 5004.

This work has made use of data from the European Space Agency (ESA) mission \textit{Gaia} (www.cosmos.esa.int/gaia), processed by the \textit{Gaia} Data Processing and Analysis Consortium (DPAC, www.cosmos.esa.int/web/gaia/dpac/consortium). Funding for the DPAC has been provided by national institutions, in particular the institutions participating in the \textit{Gaia} Multilateral Agreement.

T. C.-G. and A. V. acknowledge the support of ASI through grant 2014-025-R.1.2015.

A. M. and and A. K.-M. acknowledge support from the Portugese Funda\,c\~ao para a Ci\^encia e a Tecnologia (FCT) through grants: SFRH/BPD/74697/2010 and the Strategic Programme UID/FIS/00099/2013.

This work was supported by the MINECO (Spanish Ministry of Economy) - FEDER through grant ESP2016-80079-C2-1-R, ESP2014-55996-C2-1-R, and MDM-2014-0369 of ICCUB (Unidad de Excelencia `Mar\'ia de Maeztu'), as well as the Juan de la Cierva program 2015.

This work has made extensive use of Topcat \citep{Taylor05}, and of NASA's Astrophysics Data System.
The figures in this paper were produced with Matplotlib \citep{Hunter07} and Healpy, a Python implementation of HEALPix \citep[][]{Gorski05}.


\bibliographystyle{astron_tristan} 
\linespread{1.5}                
\bibliography{biblio}

\end{document}